\documentclass[%
 reprint,
superscriptaddress,
nofootinbib,
 amsmath,amssymb,
 aps,
 pra,
 natbib
]{revtex4-2}

\usepackage{graphicx}
\usepackage{dcolumn}
\usepackage{bm}
\usepackage{hyperref}
\usepackage{siunitx}
\usepackage[caption=false]{subfig}
\usepackage{booktabs}
\usepackage[export]{adjustbox}
\usepackage{multirow}
\usepackage{physics}
\usepackage{accents}
\usepackage{color}
\usepackage{xcolor}

\usepackage{soul}
\usepackage{ifthen}

\newboolean{showcuts}
\setboolean{showcuts}{false} 

\begin{document}

\title{Radial Fast Entangling Gates Under Micromotion in Trapped-Ion Quantum Computers}

\def\ANU{Department of Quantum Science and Technology, The Australian National University, Canberra, ACT 2601, Australia}
\def\IonQ{IonQ, Inc., College Park, MD, USA}

\def\Innsbruck{Institut für Experimentalphysik, Universität Innsbruck, Technikerstra{\ss}e 25a, 6020 Innsbruck, Austria}

\author{Phoebe Grosser}
 \affiliation{\ANU}\affiliation{\Innsbruck}%
\author{Monica Gutierrez Galan}
 \affiliation{\IonQ}
\author{Isabelle Savill-Brown}
 \affiliation{\ANU}%
   \author{Alexander K. Ratcliffe}
 \affiliation{\IonQ}%
  \author{Haonan Liu}
 \affiliation{\IonQ}%
  \author{Varun D. Vaidya}
 \affiliation{\IonQ}%
  \author{Simon A. Haine}
 \affiliation{\ANU}%
  \author{C. Ricardo Viteri}
 \affiliation{\IonQ}%
\author{Joseph J. Hope}
 \affiliation{\ANU}%
\author{Zain Mehdi}
 \email{zain.mehdi@anu.edu.au}%
 \affiliation{\ANU}%

\date{\today}

\begin{abstract}
Micromotion in radio-frequency ion traps is generally considered detrimental for quantum logic gates, and is typically minimized in state-of-the-art experiments. However, as a deterministic effect, it can be incorporated into quantum control frameworks aimed at designing high-fidelity quantum logic controls. In this work, we demonstrate that micromotion can be beneficial to the design of fast entangling gates utilizing the radial modes of a two-ion crystal, particularly in the sub-trap-period regime where high-fidelity control sequences are identified with operation times ranging from hundreds of nanoseconds to microseconds. Through analysis of select fast gate solutions, we uncover the physical origin of micromotion enhancement and further study the induced gate error under experimental noises and control imperfections. This analysis establishes the feasibility of realising high-fidelity entangling gates in hundreds of nanoseconds using the micromotion-sensitive radial modes of trapped-ion crystals. 
\end{abstract}

\maketitle

\section{Introduction}
Trapped atomic ions have a number of advantages for quantum information processing~\cite{Cirac1995a,Bruzewicz2019,Drmota2024}: exceptional coherence times of qubits encoded in electronic states~\cite{Langer2005,Harty2014,Wang2021a}, high-fidelity quantum logic operations~\cite{Harty2014,Ballance2016,Srinivas2021a,Ryan-Anderson2024,Moses2023d,Leu2023,Cai2023,Weber2024}, and strong qubit connectivity~\cite{Debnath2016,Wright2019,Hou2024,Kielpinski2002,Pino2021,Moses2023d}. State-of-the-art trapped-ion processors rely on radio-frequency (RF) Paul traps~\cite{Paul1990a}, where a strong AC driving field induces a deterministic modulation of the ion-crystal's harmonic motion, known as intrinsic micromotion~\cite{Leibfried2003b}. Micromotion is typically neglected in the design of conventional quantum logic gates, such as the M\o{}lmer-S\o{}renson entangling mechanism~\cite{Sorensen2000}, which operate on adiabatic timescales with respect to the trapping fields~\cite{Bruzewicz2019}. However, the effects of micromotion become increasingly important as gate speeds are pushed towards the non-adiabatic regime, where the control fields are adjusted at timescales comparable to or faster than the RF drive~\cite{Ratcliffe2020}. 

Non-adiabatic `fast' entangling gates use high-power laser pulses to implement two-qubit operations on timescales comparable to the trapping period~\cite{Garcia-Ripoll2005c,Steane2014}. This can be achieved either through amplitude modulation of the driving laser within the Lamb-Dicke regime~\cite{Schafer2018a}, or through sequences of state-dependent momentum kicks (SDKs) delivered by ultrafast $\pi$-pulses~\cite{Garcia-Ripoll2003c}. The latter is limited by the fidelity of those SDKs~\cite{Wong-Campos2017a} which are more experimentally challenging than MHz modulation. However, SDK-based fast gates with machine-designed pulse sequences~\cite{Gale2020} have a number of advantages: they support temperatures beyond the Lamb-Dicke regime~\cite{Taylor2017b}, scale to larger ion crystals with all-to-all connectivity~\cite{Mehdi2021e,Savill-Brown2025,Savill-Brown2025b}, operate between ions in neighbouring traps~\cite{Ratcliffe2018,Mehdi2020b,Wu2020b}, and support mixed-species ion pairings~\cite{Mehdi2025}.

The purpose of this work is to explore how micromotion can be exploited in the design of high-fidelity SDK-based fast gates utilizing the radial modes of a trapped-ion crystal. A study by Ratcliffe \textit{et al.}~\cite{Ratcliffe2020} demonstrated the existence of regimes where micromotion can improve the speed of two-qubit fast gates by locking laser pulses to a specific phase of the RF drive, with larger micromotion amplitudes leading to greater potential benefit. However, this phase-locking also places the strong constraint that the RF frequency must be a multiple of the laser repetition rate. Furthermore, this restriction ignores all gate solutions that leverage the full temporal structure of the RF drive, which opens up many new control degrees of freedom.

This work extends the findings of Ratcliffe \textit{et al.}~\cite{Ratcliffe2020} to a phase-unlocked regime, where SDKs are no longer constrained to arrive at a fixed point in the RF cycle. Using numerical optimization over high-dimensional pulse sequences~\cite{Gale2020}, we identify high-fidelity entangling gate solutions with operation times comparable to and faster than a single radial trapping period across a wide range of micromotion environments. We find that larger amplitude micromotion can improve both gate speed and fidelity; the latter by up to two orders of magnitude. We analyse the physical origin of this enhancement in terms of the deviation from the secular (micromotion-free) limit, and study the susceptibility of these gates to experimental error sources, finding that fidelities exceeding $0.9$ may be achieved using previously demonstrated SDK fidelities, and that gate errors approaching $10^{-5}$ are possible with high-fidelity SDKs~\cite{CW_SDK}. Our results establish the viability of achieving MHz-rate entangling gates in environments with significant micromotion, such as the radial modes of a Paul trap.

\section{Theory of fast gates under micromotion}

In this section we outline the SDK-based fast gate formalism in the presence of micromotion. This extends the original fast gate formalism developed in Ref.~\cite{Garcia-Ripoll2003c} to incorporate the full time-dependent dynamics of the RF trap (see Appendix~\ref{sec:BackgroundTheory}). The key result of this section is a set of condition equations for high-fidelity entanglement [Equations~\eqref{eq:Phase_ConditionEquations}-\eqref{eq:Y_ConditionEquations}], which highlight micromotion-induced corrections to the gate dynamics. These results are valid for arbitrary length chains, though we will focus on the $N=2$ case for the remainder of this work.

\begin{figure*} \includegraphics[width=2\columnwidth]{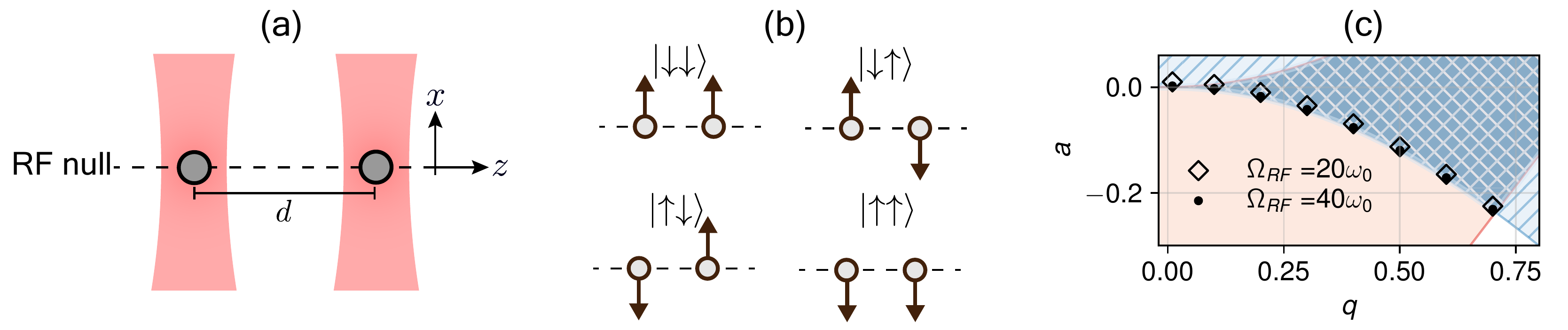}
\caption{\label{fig:Schematic}
Impulsive fast two-qubit gates using the transverse (radial) modes of a two-ion crystal. 
(a) Ions are subjected to SDKs using focussed optical beams (shown in red) transverse to the null of the RF trap (dashed line). 
(b) Addressing both ions with SDKs excites the in-phase (`common-motional') mode for same-spin two-qubit states ($\ket{\downarrow\downarrow}$ and $\ket{\uparrow\uparrow}$), and the out-of-phase (`stretch') mode for different-spin two-qubit states ($\ket{\downarrow\uparrow}$ and $\ket{\uparrow\downarrow}$).(c) First-order stability region for the trapping parameters $(a,q)$ along the $x$-axis (hatched region shaded in blue) and the $y$-axis (red shaded region). The overlap between the two regions corresponds to stable trapping in all dimensions (cross-hatched region). Trapping parameter values along the $x$-axis chosen to maintain the same trapping potential in the secular limit indicated with dots (diamonds) for an RF drive of $40\omega_0$ ($20\omega_0$), where $\omega_0$ is the secular frequency of the RF trap. }
\end{figure*}

\subsection{State-dependent kicks}
\label{sec:SDK}
Fast gates rely on the use of broadband (picosecond to nanosecond) $\pi$-pulses that impulsively transfer internal-state-dependent momentum to each of the addressed ions as illustrated in Fig.~\ref{fig:Schematic}. Each of these SDKs can be realised by addressing each ion with a laser pulse of area $\pi$, resonant with the qubit frequency with effective wavevector $k$. A $\pi$-pulse incident on the $j$-th ion is described by the unitary:
\begin{equation}
\hat{U}_{k, \phi_L}^j(t) = \hat{\sigma}_- e^{i(k\hat{x}^{(j)}(t)+\phi_L)} +\hat{\sigma}_+ e^{-i(k\hat{x}^{(j)}(t)+\phi_L)} \, ,
\end{equation}
where $\phi_L$ is the phase of the laser electric field and $\hat{x}^{(j)}(t)$ describes the deviation of the $j$-th ion's position from the RF null (see Fig.~\ref{fig:Schematic}(a)).Here we are working in an interaction picture with respect to the motional dynamics of the ion crystal in a time-dependent RF potential (see Appendix \ref{sec:BackgroundTheory}):
\begin{align}
	V_{\rm RF}(x,t) = \frac{m \Omega_{\rm RF}^2}{8}\left[a_x-2q_x\cos(\Omega_{\rm RF} t +\phi_{\rm RF}) \right]x^2 \,,
\end{align}
characterised by the dimensionless parameters $a_x$ and $q_x$.

In this work we consider a laser-phase-insensitive `paired' SDK protocol, where each SDK is composed of a pair of counter-propagating $\pi$-pulses incident on each ion (e.g. by retroreflecting each pulse~\cite{Bentley2013,Putnam2024}). If the phase of the two pulses in each pair are the same (which can be ensured by splitting them from a $2\pi$ parent pulse), the laser-phase dependence of the paired SDK vanishes:
\begin{align}
\hat{U}_{\rm{SDK}}(k,t) &= \prod_{j=A,B} \hat{U}_{-k, \phi}^j(t)\hat{U}_{k, \phi}^j(t) \\
& = \exp \left\{2i  k (\hat{x}^A(t) \hat{\sigma}_z^A + \hat{x}^B(t) \hat{\sigma}_z^B) \right\} \, . \label{eq:SDK_PositionBasis}
\end{align}
The action of this unitary on the two-qubit system is illustrated in Fig.~\ref{fig:Schematic}(b). We will assume a paired SDK scheme for the remainder of this work, though the extension to unpaired $\pi$-pulses is straightforward~\cite{Wong-Campos2017a,Savill-Brown2025b}.

\subsection{The fast gate unitary}
\label{sec:FastGateUnitary}
A fast two-qubit gate can be built up from a sequence of $\mathcal{N}$ SDKs arriving at times $\{t_1,t_2,\dots,t_\mathcal{N}\}$, separated by periods of free motional evolution. Working in the interaction picture with respect to the motional Hamiltonian of the system, the gate unitary can be expressed as a time-ordered product of SDKs:
\begin{align}
    \hat{U}_{\rm G}  &= \mathcal{T}\prod_{j=1}^\mathcal{N}\hat{U}_{\rm{SDK}}(z_j k,t_j) \,, 
\end{align}
with $\mathcal{T}$ denoting the time-ordering operator. Here we have allowed for switching the direction of each SDK between $+k$ and $-k$ by the introduction of the parameter $z_j = \pm 1$ for the $j$-th SDK. Applying the Baker-Campbell-Hausdorff lemma, $ \hat{U}_{\rm G}$ may be expressed as the product of two commuting unitaries, i.e. $\hat{U}_{\rm G}=\hat{U}_2\hat{U}_1$, where the first unitary describes an effective spin-spin interaction between the addressed qubits:
\begin{align}
    \label{eq:SpinSpinUnitary} \hat{U}_1 = \exp\left\{i\Theta  \hat{\sigma}_z^{(A)}\otimes \hat{\sigma}_z^{(B)} \right\} \,,
\end{align}
and the second describes residual spin-motional coupling at the end of the gate operation:
\begin{align}
   \hat{U}_2 = \bigotimes_\alpha \hat{D}_\alpha\left\{\Delta \xi_\alpha  \sum_{j=A,B}b_\alpha^{(j)} \hat{\sigma}_z^{(j)} \right\} \,,
\end{align}
where $\Delta\xi_\alpha$ is the amplitude of residual spin-dependent displacement of the $\alpha$-th motional mode at the end of the gate operation, and $b_\alpha^{(j)}$ is the normalized coupling of the $j$-th ion to the $\alpha$-th motional mode. For the two-ion system, $b_{\rm CM}^{(1)}=b_{\rm CM}^{(2)}=1/\sqrt{2}$ for the center-of-mass (CM) mode and $b_{\rm SR}^{(1)}=-b_{\rm SR}^{(2)}=1/\sqrt{2}$ for the stretch (SR) mode.

Provided that the motional states are decoupled from the internal states at the conclusion of the gate operation (i.e. $\Delta\xi_\alpha =0$), the gate unitary takes the form of a ZZ entangling operation [$U_{\rm ZZ}(\theta)=e^{i\theta\hat{\sigma}_z^{(A)}\otimes \hat{\sigma}_z^{(B)} }$]. Maximum entanglement is achieved for $\Theta_{\rm targ} =\pm \pi/4$. The key challenge to designing a high-fidelity operation is finding a pulse sequence $\{z_k,t_k\}$ that satisfies the maximally-entangling condition, $\Theta =\pm \pi/4$, and restores all motional modes [$\Delta\xi_\alpha=0$]. We describe a machine-design approach to address this problem in the following section. \par

The timings and directions of the SDK sequence enter the gate unitary through the quantities $\Theta $ and $\Delta\xi_\alpha $, which respectively describe the relative phases between the eigenstates of $\hat{\sigma}_z^{(A)}\otimes \hat{\sigma}_z^{(B)}$ and the residual spin-dependent displacement amplitude of the $\alpha$-th motion mode. Expressing the residual displacement in terms of the dimensionless mode quadratures, i.e. $\Delta X_\alpha = [\Delta\xi_\alpha u_\alpha^*(t_{\rm g})+\Delta\xi_\alpha^* u_\alpha(t_{\rm g})]/\sqrt{2}$ and $\omega_\alpha\Delta Y_\alpha = [\Delta\xi_\alpha \dot{u}_\alpha^*(t_{\rm g})+\Delta\xi_\alpha^* \dot{u}_\alpha(t_{\rm g})]/\sqrt{2}$, the following expressions for these quantities are derived in Appendix \ref{app:FGDerivation}:
\begin{widetext}
\begin{subequations}
\label{eq:ConditionEquations}
    \begin{align}
        \Theta &= 8\sum_\alpha \eta_\alpha^2 b_\alpha^{(A)}b_\alpha^{(B)}\sum_{n=1}^\mathcal{N}\sum_{m=2}^{n-1}\frac{z_nz_m}{\rho_\alpha(t_m)}\left[\sin(\omega_\alpha\Delta t_{nm})\mu_\alpha^{(c)}(t_n,t_m) +\cos(\omega_\alpha\Delta t_{nm})\mu_\alpha^{(s)}(t_n,t_m)\right]\, \label{eq:Phase_ConditionEquations}\\
        \Delta X_\alpha &= \sqrt{2}\eta_\alpha  \sum_{j=1}^\mathcal{N} \frac{z_j}{\rho_\alpha(t_j)}\left[\sin(\omega_\alpha[t_{\rm g}-t_j])\mu_\alpha^{(c)}(t_{\rm g},t_j) +\cos(\omega_\alpha[t_{\rm g}-t_j])\mu_\alpha^{(s)}(t_{\rm g},t_j)\right] \label{eq:X_ConditionEquations}\\
         \Delta Y_\alpha &= \sqrt{2}\eta_\alpha  \sum_{j=1}^\mathcal{N} \frac{z_j}{\rho_\alpha(t_j)}\left[\sin(\omega_\alpha[t_{\rm g}-t_j])\kappa_\alpha^{(c)}(t_{\rm g},t_j) +\cos(\omega_\alpha[t_{\rm g}-t_j])\kappa_\alpha^{(s)}(t_{\rm g},t_j)\right] \label{eq:Y_ConditionEquations}
    \end{align}
\end{subequations}
\end{widetext}
where $\Delta t_{nm}\equiv t_n-t_m$ and we have defined $\rho_\alpha(t_j) = 1+(f_{\alpha,C}(t_j)\dot{f}_{\alpha,S}(t_j)-f_{\alpha,S}(t_j)\dot{f}_{\alpha,C}(t_j))/\omega_\alpha$ in terms of the functions $f_{\alpha,C}(t)$ and $f_{\alpha,S}(t)$ which describe high-frequency oscillations at harmonics of the RF drive (\textit{c.f.}~Eq.~\eqref{eq:MM_modefunctions} in Appendix~\ref{sec:BackgroundTheory}):
\begin{subequations}
    \begin{align}
	f_{\alpha,S}(t) &\equiv \sum_{n}C_n\sin(n\Omega_{\rm RF} t+n\phi_{\rm RF}) \,, \\
		f_{\alpha,C}(t) &\equiv \sum_{n}C_n\cos(n\Omega_{\rm RF} t+n\phi_{\rm RF})  \,.
\end{align}
\end{subequations}
The condition equations given in Eq.~\eqref{eq:ConditionEquations} take a similar form to the result for the case of purely secular motion~\cite{Garcia-Ripoll2003c,Ratcliffe2020} with corrections due to RF micromotion encoded in the dimensionless variables:
\begin{subequations}
\label{eq:TensorsDef_mu}
\begin{align}
	\mu_\alpha^{(s)}(t,t') &\equiv f_{\alpha,C}(t')f_{\alpha,S}(t)-f_{\alpha,C}(t)f_{\alpha,S}(t') \,, \\
	\mu_\alpha^{(c)}(t,t') &\equiv f_{\alpha,C}(t)f_{\alpha,C}(t')+f_{\alpha,S}(t)f_{\alpha,S}(t') \,.
   \end{align}
\end{subequations}
These variables, which modify the result for the final position quadrature of each motional mode ($\Delta X_\alpha$), are interpreted as sine-like and cosine-like, respectively, due to their (anti)-symmetry properties: $\mu_\alpha^{(s)}(t,t')=-\mu_\alpha^{(s)}(t',t)$ and $\mu_\alpha^{(c)}(t,t')=\mu_\alpha^{(c)}(t',t)$. Similarly, the dimensionless variables
\begin{subequations}
\label{eq:TensorsDef_kappa}
    \begin{align}
        \kappa_\alpha^{(s)}(t,t') &= \partial_t \mu_\alpha^{(c)}(t,t')/\omega_\alpha- \mu_\alpha^{(s)}(t,t')\,, \\
   \kappa_\alpha^{(c)}(t,t') &= \partial_t \mu_\alpha^{(s)}(t,t')/\omega_\alpha+ \mu_\alpha^{(c)}(t,t')
    \end{align}
\end{subequations}
encode the modification of the final momentum quadrature of each motional mode ($\Delta Y_\alpha$) due to RF micromotion. These are related to the derivatives of $\mu^{c,s}(t,t')$ due to the relationship between the mode quadratures, $Y_\alpha = \dot{X}_\alpha/\omega_\alpha$ (see Eq.~\eqref{eq:app:HeisenbergXY_Dimensionless} in Appendix~\ref{app:FGDerivation}). \par 

The deviation of these variables from their secular limits (e.g. $\mu_\alpha^{(c)}(t,t') \approx 1$ and $\mu_\alpha^{(s)}(t,t')\approx 0$ to lowest order in $q_x^2\ll 1$) can be tuned by adjusting the timing of each SDK with respect to the phase of the RF cycle. For example, locking all SDKs to the same point in the RF cycle (i.e. enforcing $\Omega_{\rm RF} t_j+\phi_{\rm RF} = \phi_0$) reduces the condition equations Eq.~\eqref{eq:ConditionEquations} to secular form with an effective rescaling of the phase accumulation rate and residual spin-dependent motion by $1/\rho_\alpha(t_j) \approx [1+q_x\cos(\phi_0)/2]^{-1}$. This approach has been explored previously~\cite{Ratcliffe2020} as a method of improving gate speed (for a fixed number of pulses), at the cost of increasing errors due to residual spin-motional coupling. In this work we take a more general approach, wherein we allow each SDK to arrive at a different phase of the RF cycle, using a machine-design approach that we will describe in the following section.

\subsection{State-averaged infidelity measure}
\label{sec:Infidelity}
In order to realise a high-fidelity maximally-entangling ZZ operation with the fast gate mechanism, the following $2N+1$ commensurability conditions must be met (for a system with $N$ motional modes):
\begin{subequations}
    \begin{align}
        \Theta &= \pm \pi/4 \,, \\
        \Delta X_\alpha &=0 \,, \\
        \Delta Y_\alpha &=0 \,.
    \end{align}
\end{subequations}
The error of a particular gate solution (i.e., a particular set of $\{z_k,t_k\}$ with $\mathcal{N}=\sum_j|z_j|$ total SDKs) can be quantified in terms of the state-averaged gate infidelity assuming that the motional degrees of freedom are initially at thermal equilibrium with temperature $T$~\cite{Ratcliffe2020, Bentley2015a}:
\begin{align}
    \label{eq:StateAvg_Inf} 1 - \mathcal{F} \approx  \frac{2}{3} \Delta \Phi^2 &+\frac{4}{3} \sum_\alpha \left[ \frac{1}{2} + \bar{n}_\alpha(T) \right] \\ \notag &\times \left[\left( b_\alpha^{(A)} \right)^2+\left( b_\alpha^{(B)} \right)^2\right] \left[ \Delta X_\alpha^2 + \Delta Y_\alpha^2 \right] \, , 
\end{align}
where $\Delta \Phi = |\Theta|-\pi/4$ is the two-qubit relative phase error and the mean phonon occupation of each mode follows the Bose-Einstein distribution, i.e. $\bar{n}_\alpha(T) = [\exp(\hbar\omega_\alpha/k_B T)-1]^{-1}$. 
Equation \eqref{eq:StateAvg_Inf} assumes small errors, which is sensible for gate solutions with fidelity close to unity. 
The expression has leading corrections of order $\mathcal{O}(\Delta \Theta^4,|\Delta X_\alpha|^3,|\Delta Y_\alpha|^3)$ (see the supplementary material of Ref.~\cite{Mehdi2025} for the all-order expression), which rapidly vanish for gate solutions with $\mathcal{F}\gtrsim 0.99$.

\subsection{Insensitivity to excess micromotion}

We briefly note that the SDK-based entangling gate mechanism is insensitive to excess micromotion -- oscillations at the RF frequency that occur when the ions are displaced from the RF null~\cite{Berkeland1998}, e.g. due to stray electric fields. This is because, as a geometric phase gate protocol~\cite{Garcia-Ripoll2005c}, the fidelity of a fast gate operation does not depend on the initial position or momentum of the ions. This can be understood by studying the spin-spin term in the exponent of the fast gate unitary, Eq.~\eqref{eq:SpinSpinUnitary},  which is non-zero due to the non-commutativity of individual SDK unitaries at different times. This cannot be mimicked by state-independent displacements of the RF null as the mapping $\hat{x}^{(i)}\rightarrow \hat{x}^{(i)}+\epsilon$ results in terms that are linear in $\hat{\sigma}_z$ in the exponent of the fast gate unitary.

\section{Optimization Methods} 
\label{sec:Methods}

We perform high-dimensional numerical optimization to search for pulse sequences that accumulate the correct entangling phase ($\Delta \Phi = 0$) and disentangle the qubit states from each motional mode (i.e. $\Delta X_\alpha = \Delta Y_\alpha = 0$). As optimisation over the unconstrained $2 \mathcal{N}$-dimensional parameter space is generally intractable, we focus our search to pulse sequences composed of short SDK `groups' separated by large periods of free evolution~\cite{Garcia-Ripoll2003c,Duan2004a,Bentley2013,Ratcliffe2018}. In this study, we employ the Generalised Pulse Group (GPG) scheme~\cite{Gale2020}, a two-stage algorithm that has been used to find high-fidelity gate solutions across experimentally-relevant regimes~\cite{Mehdi2020b,Mehdi2021e,Mehdi2025,Savill-Brown2025,Savill-Brown2025b}.

In the GPG scheme, the pulse sequence is divided into $M$ SDK `groups', each delivering an integer number of momentum kicks ($z_j$) to the addressed ions. The optimization algorithm has two stages. In the first stage, a stochastic global search is performed over the kick magnitudes $\{ z_1,\dots,z_M \}$ while enforcing uniform timings for the pulse groups -- the $j$-th group is centered around time $t_j=t_g(j-1)/(M-1)$. The second stage is a local gradient-descent optimization of the group timings $\{t_j\}$. We include the effect of a finite laser repetition rate in the second stage by enforcing a minimum spacing of $1/f_{\rm{rep}}$ between consecutive SDKs, where $f_{\rm{rep}}$ is the SDK repetition rate.

Further details of our optimization methodology are outlined in Refs.~\cite{Gale2020,Mehdi2025,Savill-Brown2025b}. We note that the GPG optimisation protocol~\cite{Gale2020} samples a subset of the full pulse sequence parameter space. Solutions presented in this work are not guaranteed to be global optima.

For the purposes of designing fast gates for two-ion systems and comparing gate searches between different micromotion environments, we keep both the RF and secular timescales of the dynamics fixed when optimising Eq.~\eqref{eq:StateAvg_Inf} for trapping geometries with different $q_x$ values. As there are many different $\{ a_\alpha, q_x \}$ values that exhibit the same secular limit $\omega_\alpha \approx \Omega_{\rm{RF}} \sqrt{a_\alpha + q_x^2/2}/2$, we choose $a_\alpha$ for a given $q_x$ self-consistently so that all micromotion environments exhibit the same secular limit. Further detail on this process is provided in Appendix \ref{app:opti-params}. Unless otherwise stated, the default parameter values utilised in this paper (in trap units set by $\omega_0 \equiv \omega_{\rm CM}$) are: 
$\Omega_{\rm{RF}} = 40\omega_0$ and $(\omega_{\rm{SR}} - \omega_{\rm{CM}})/\omega_{\rm{CM}} = -1.4 \times 10^{-2}$~\cite{Ratcliffe2020}. The corresponding trapping parameters are illustrated in Fig.~\ref{fig:Schematic}(c). For all calculations we assume a Lamb-Dicke parameter of $\eta \approx 0.15$, which is consistent with the use of Ba$^+$ trapped-ion qubits controlled by counter-propagating $532$nm Raman lasers and a radial trapping frequency of $2\pi\times 1$MHz. All timescales in this work will be reported in terms of trapping periods, i.e. units of $2\pi/\omega_0$ where the $\omega_0 \equiv \omega_{\rm CM} = \beta_{\rm CM}\Omega_{\rm RF}/2$ is the centre-of-mass oscillation frequency; henceforth referred to as the `secular trapping frequency'.

\section{Results: Micromotion-enhanced fast gate solutions} 
\label{sec:Results}
\begin{figure*} \includegraphics[width=2\columnwidth]{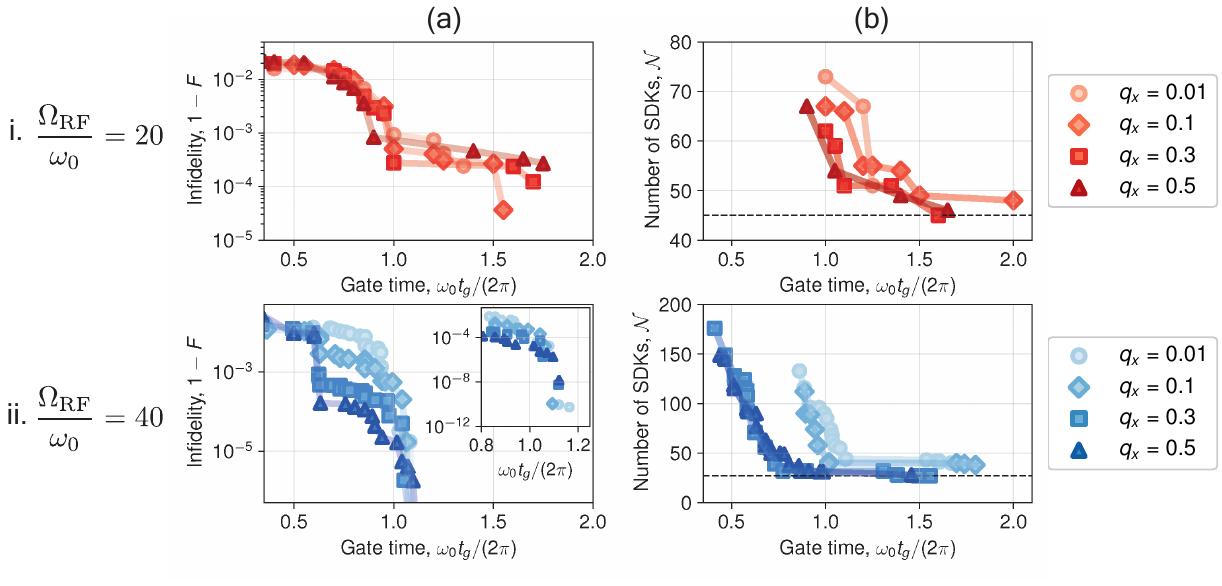}
\caption{Micromotion enhancement of radial fast gates in a RF trap with relative mode splitting $\chi=-1.4\times 10^{-2}$ and RF drive frequency i. $\Omega_{\rm RF}=20\omega_0$ and ii. $\Omega_{\rm RF}=40\omega_0$.
(a) Infidelities of generalised fast gate solutions as a function of gate time in trap periods. For fast gates with gate times between $0.6 \le \omega_0 t_{\rm g} / (2 \pi) \le 1.0$ trap periods, gate errors are suppressed in environments with larger micromotion amplitudes by up to two orders of magnitude, for solutions with less than $80$ SDKs. 
(b) Number of SDKs per gate ($\mathcal{N}$) for gate solutions with stage-averaged fidelities exceeding 99.9\%, with horizontal dashed lines indicating the minimum number of SDKs to implement a maximally entangling gate in less than two trapping periods.}
\label{fig:MMEnhancement_InfRR} 
\end{figure*}

We begin by exploring the role of micromotion in the trade-off between gate speed and fidelity. We perform scans over the target gate time, focusing specifically on the sub-trap-period regime, $\omega_0 t_{\rm g}/(2\pi)\lesssim 1$. Multiple solutions are often found for the same target gate time; we choose the `optimal' solutions by identifying solutions with the fewest number of SDKs for a given fidelity threshold -- we take this to be $99.9\%$ in the context of the present work. For simplicity we assume the limit of infinite SDK repetition rate, which places an upper bound on gate speed and fidelity. The role of the finite SDK repetition rate will be studied in the following section.

\subsection{Trade-off between gate speed, fidelity, and number of SDKs}
Figure \ref{fig:MMEnhancement_InfRR} presents a set of fast gate solutions optimised using the GPG scheme for two different RF drive frequencies, $\Omega_{\rm RF}=20\omega_0$ and $\Omega_{\rm RF}=40\omega_0$. 
Fig.~\ref{fig:MMEnhancement_InfRR}(a) illustrates that the trade-off between gate speed and fidelity depends on the ratio of the RF drive frequency to the secular trapping frequency.
For the lower frequency RF drive, $\Omega_{\rm RF}=20\omega_0$, we find high fidelity solutions ($F>99.9\%$) for gate durations longer than a single trapping period across all micromotion amplitudes ($q_x$), with the best achievable fidelities ranging between $99.9\%$ and $99.99\%$ for gate durations between one and two trapping periods. 
In the sub-trap-period regime, the state-averaged fidelity decays to below $99\%$ as the gate duration approaches half a trapping period.

Fig.~\ref{fig:MMEnhancement_InfRR}(a) demonstrates qualitatively distinct trends for the higher frequency RF drive, $\Omega_{\rm RF}=40\omega_0$. In the sub-trap-period regime, $0.5\lesssim\omega_0 t_{\rm g}/(2\pi)\lesssim 1$, we observe that larger micromotion amplitudes ($q_x$) enable higher fidelities to be achieved. In particular, we observe suppression of gate errors (due to imperfect phase accumulation and motional restoration) by three orders of magnitude for gates optimized in a high micromotion environment of $q_x=0.5$ as compared to a minimal micromotion environment of $q_x=0.01$. Furthermore, the inset in Fig.~\ref{fig:MMEnhancement_InfRR}(a.ii) shows that very high fidelities ($1-F\ll 10^{-8}$) are achievable for gate times slightly longer than a single trap period, across all micromotion environments. This is to be contrasted with the plateauing of gate fidelity between $99.9\%$ and $99.99\%$ when optimized for a lower RF drive frequency as shown in Fig.~\ref{fig:MMEnhancement_InfRR}(a.i). These results illustrate the importance of a relatively high RF drive frequency to the \textit{enhancement} of fast two-qubit gates due to micromotion. This has a clear physical interpretation: in order for micromotion to provide additional control over the state-dependent motion of the ions, there must be a sufficient number of RF cycles during the gate. The results presented in Fig.~\ref{fig:MMEnhancement_InfRR} suggest the minimum number of RF cycles per gate  required for high-fidelity solutions is approx. $20$ for the specific trapping geometry and ion species considered in this work.

We further analyse the micromotion enhancement of fast gate solutions in Figure \ref{fig:MMEnhancement_InfRR}(b), where we study the trade-off between gate duration and number of SDKs required per gate ($\mathcal{N}\equiv \sum_j |z_j|$) for solutions with state-averaged fidelity above $99.9\%$. The latter is an important metric as the contribution of pulse imperfections to the overall two-qubit gate error is expected to scale with $\mathcal{N}$~\cite{Bentley2016a,Taylor2017b,Gale2020,Savill-Brown2025}. For the case of the higher RF drive, $\Omega_{\rm RF}=40\omega_0$, Fig.~\ref{fig:MMEnhancement_InfRR}(b.ii) illustrates a significant reduction in the number of SDKs required for high-fidelity gate solutions at a particular gate time as the amplitude of micromotion ($q_x$) is increased in magnitude. This enables faster gate solutions with fewer pulses. For example, with $\mathcal{N}\leq 40$ SDKs, the fastest gate solution with fidelity above $99.9\%$ is $1.8 \times 2\pi/\omega_0$ for $q_x = 0.01$ as compared to $0.8\times 2\pi/\omega_0$ for $q_x=0.5$. This feature is not observed for the lower RF drive as shown in Fig.~\ref{fig:MMEnhancement_InfRR}(a.ii), which further emphasises the importance of high RF to secular frequency ratios to the micromotion enhancement of high-fidelity gate solutions~\cite{Ratcliffe2020}.

\begin{figure*} \includegraphics[width=1.88\columnwidth]{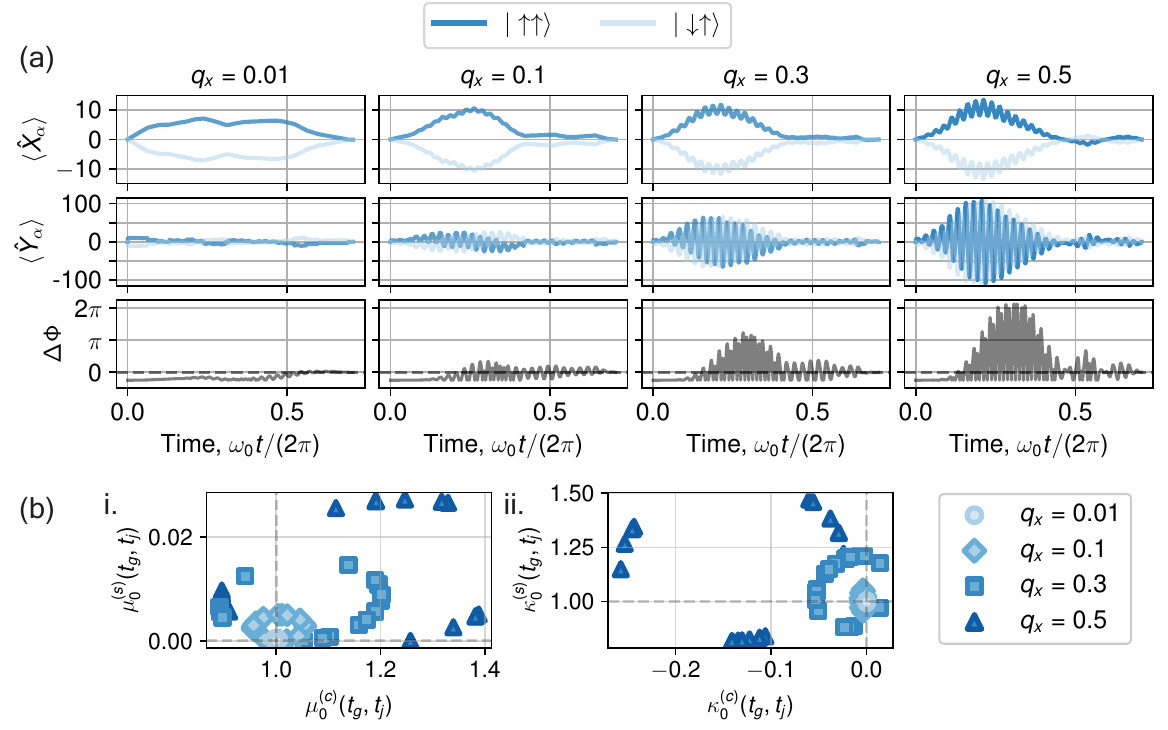}
\caption{Analysis of select gate solutions from Fig.~\ref{fig:MMEnhancement_InfRR}ii. with sub-trap-period duration $\omega_0 t_{\rm g} /(2 \pi) \approx 0.7$, illustrating micromotion enhancement. (a) Motional dynamics of different two-qubit states in terms of the dimensionless position and momentum of each mode, and the difference between the accumulated two-qubit phase $\Theta$ and its target value, $\Delta \Phi = \Theta - \pi/4$. The centre-of-mass mode couples solely to same-spin two-qubit states $\{\ket{\uparrow\uparrow},\ket{\downarrow\downarrow}\}$, and the stretch mode couples solely to the anti-aligned spin states $\{\ket{\downarrow\uparrow},\ket{\uparrow\downarrow}\}$. (b) i. The vectors $\mu^{(s)}_0(t_{\rm g},t_j)$ and $\mu^{(c)}_0(t_{\rm g},t_j)$ are plotted for all SDK timings $\{t_j\}$, and can be interpreted as high-frequency corrections to the time-dependent ion positions as compared to the secular limit (indicated by the dashed grey lines). ii. Similarly, the vectors $\kappa^{(s)}_0(t_{\rm g},t_j)$ and $\kappa^{(c)}_0(t_{\rm g},t_j)$ are the micromotion corrections to the time-dependent momentum of the ions, \emph{c.f} Eq.~\eqref{eq:ConditionEquations}.}
\label{fig:TuningParams_MMEnhancement_InfRR} 
\end{figure*}
\subsection{Analysis of example gate solutions}

To further analyze the enhancement of fast gate solutions due to micromotion, we consider a set of example gate solutions that all share the same approximate duration of $t_{\rm g}=0.7\times2\pi/\omega_0$ trapping periods. In Figure \ref{fig:TuningParams_MMEnhancement_InfRR}(a) we plot the motional trajectories of the two-ion system in terms of the expectation value of the (dimensionless) position and momentum of each motional mode, and the two-qubit phase error $\Delta \Phi = |\Theta|-\pi/4$ for an initial vacuum state. In this case, an ideal gate requires $\langle \hat{X}_\alpha \rangle = \langle \hat{Y}_\alpha \rangle = \Delta \Phi =0$ at the end of the gate (i.e. immediately after the final SDK at $t=t_{\rm g}$). By symmetry, we need only to consider the two-qubit basis states $\ket{\uparrow\uparrow}$ and $\ket{\downarrow\uparrow}$ which respectively couple to the center-of-mass and stretch modes. Due to the small mode splitting between the centre-of-mass and stretch modes ($\chi = -1.4\times 10^{-2}$), the two trajectories are nearly identical.

We first consider the lowest micromotion environment, $q_x=0.01$, which closely resembles purely secular motion during the gate operation with the relative two-qubit phase increasing roughly monotonically towards $\pi/4$ (the maximally-entangling value). This indicates that the speed of the gate in this regime is limited by how fast phase can be accumulated. In contrast, we see for the higher micromotion environments that the relative phase overshoots the target value of $\pi/4$, and oscillates rapidly at the frequency of the RF drive until the end of the gate where it returns to the target value. The magnitude of the overshoot increases with micromotion amplitude, reaching $2\pi$ for $q_x=0.5$. The increased rate of phase accumulation can be attributed to the increased maximum amplitude of qubit-state-dependent displacement during the gate with increasing micromotion amplitude, as can clearly be seen in Fig.~\ref{fig:TuningParams_MMEnhancement_InfRR}(a). This provides physical insight into the mechanism behind micromotion enhancement of gate fidelity observed in Figure~\ref{fig:MMEnhancement_InfRR}: as the micromotion amplitude is increased, the accumulation of relative phase within the gate duration no longer limits the speed of the gate. As a result, the control scheme has more resources (SDKs) available to fine-tune the motional trajectories and suppress residual spin-motional entanglement (i.e. improve motional restoration). This is consistent with the results shown in Fig.~\ref{fig:MMEnhancement_InfRR} for gate durations longer than a single trapping period, where this micromotion enhancement vanishes. We interpret this as the rate of phase accumulation no longer being limited by the finite gate duration.

We further analyse the micromotion enhancement of the sub-trap-period fast gate solutions in Fig.~\ref{fig:TuningParams_MMEnhancement_InfRR}(b.i), where we plot the vectors $\mu^{(s)}_0(t_{\rm g},t_j)$ and $\mu^{(c)}_0(t_{\rm g},t_j)$ for all SDK timings $\{t_j\}$, which quantify the deviation of the final centre-of-mass position of the two-ion crystal from the case of purely secular (low frequency) dynamics -- \textit{c.f.} Eq.~\eqref{eq:ConditionEquations}. In the secular limit, $\mu^{(s)}_0(t_{\rm g},t_j)\rightarrow 0$ and $\mu^{(c)}_0(t_{\rm g},t_j)\rightarrow 1$. Similarly, Fig.~\ref{fig:TuningParams_MMEnhancement_InfRR}(b.ii) plots the vectors $\kappa^{(s)}_0(t_{\rm g},t_j)$ and $\kappa^{(c)}_0(t_{\rm g},t_j)$, which encode the micromotion corrections to the final momentum of each motional mode, with secular limits $\kappa^{(s)}_0(t_{\rm g},t_j)\rightarrow 1$ and $\kappa^{(c)}_0(t_{\rm g},t_j)\rightarrow 0$. We refer to these vectors collectively as `micromotion tuning parameters', as they allow the motional trajectories to be fine-tuned by adjustment of the SDK arrival times within the RF cycle.

Fig.~\ref{fig:TuningParams_MMEnhancement_InfRR}(b) demonstrates that the $q_x=0.01$ gate solution is effectively secular, in the sense that the position and momentum tuning parameters are well approximated by their respective secular limits. We emphasise that, despite micromotion not playing a significant role in the dynamics for $q_x=0.01$, it still needs to be taken into account in both the gate search and final evaluation of the gate fidelity. For larger micromotion amplitudes, $q_x=0.1-0.5$, Fig.~\ref{fig:TuningParams_MMEnhancement_InfRR}(b) illustrates that the tuning parameters (sampled at each SDK) span a range of values, with the magnitude of deviation from the secular limit increasing monotonically with $q_x$. This behaviour is in contrast with the phase-locked scheme of Ref.~\cite{Ratcliffe2020}, wherein the tuning parameters are made time independent by locking the SDK arrival times to a particular phase in the RF cycle. This concretely demonstrates the advantage of a phase-unlocked scheme, as considered in this work, to fully leverage micromotion towards high-fidelity and high-speed entangling gates in trapped-ion systems.

\subsection{Finite laser repetition rate \label{sec:RepRate}}
\begin{figure*} 
\includegraphics[width=1.9\columnwidth]{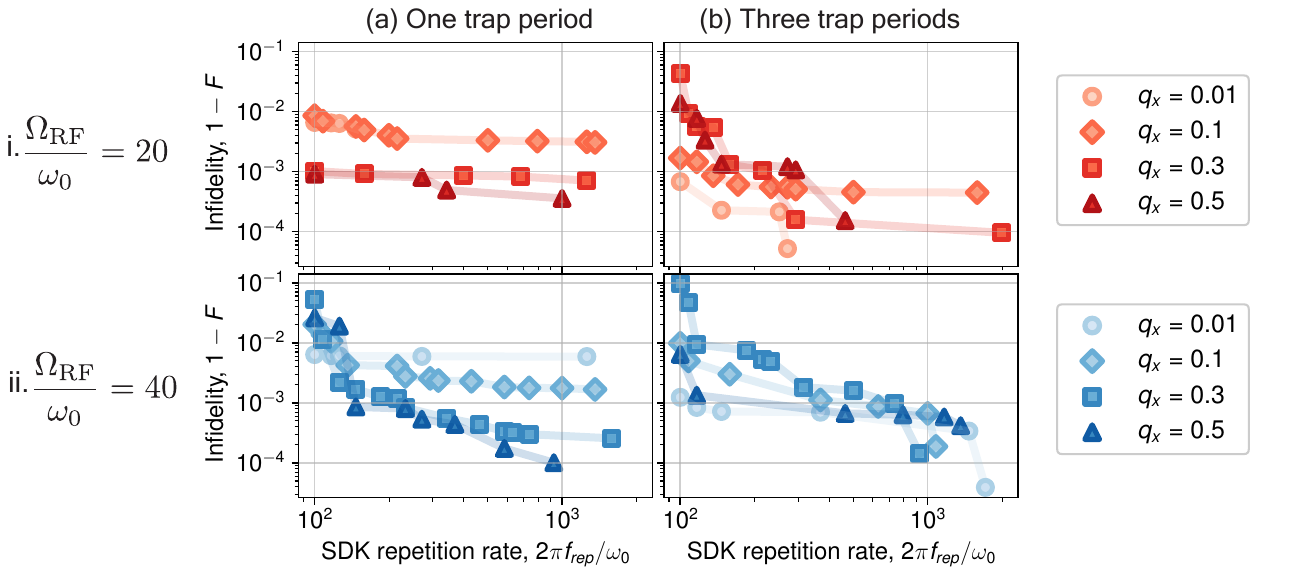}
\caption{Infidelities of fast gates designed with a finite repetition rate for RF frequencies of i.~$\Omega_{\rm{RF}} = 20\omega_0$ and ii.~$40\omega_0$. (a) Fast gates of duration $\omega_0 t_{\rm g} / (2 \pi) = 1.0$ trap periods that are optimised and evaluated with a finite repetition rate $f_{\rm rep}$. For micromotion amplitudes of $q_x \ge 0.3$, $2\pi f_{\rm{rep}}/\omega_0= $ 100 -- 300 is sufficient to achieve 99.9\%-fidelity fast gates for most environments. (b) Fast gate solutions optimized for a target duration $\omega_0 t_{\rm g} / (2 \pi) = 3.0$ trap periods.
}
\label{fig:Rep_Rate_Comp}
\end{figure*}

The above analysis has assumed the limit of infinite SDK repetition rate, in order to enable large momentum transfer to the ions by concatenating multiple SDKs with very short delays (picosecond) that are effectively instantaneous with respect to the ion dynamics. While this can be achieved using nested delay lines~\cite{Bentley2013,Wong-Campos2017a}, this adds additional experimental complexity which may not be amenable to scalable trapped-ion architectures. As an alternative, we directly incorporate the finite repetition rate of the laser ($f_{\rm rep}$) in our gate design, by imposing the constraint that consecutive SDKs must be temporally separated by at least $1/f_{\rm rep}$ in the second stage of the GPG optimization method~\cite{Gale2020,Mehdi2025}. This results in a coarse-graining of the achievable temporal control of SDKs within the RF cycle, the effect of which depends both on the ratio of the RF drive frequency to the secular trapping frequency, $\Omega_{\rm RF}/\omega_0$, and the gate time in (secular) trap periods, $\omega_0 t_{\rm g}/(2\pi)$. 

We present results for gate solutions with durations of one and three centre-of-mass oscillation periods in Figure \ref{fig:Rep_Rate_Comp}. The former case corresponds to a region in Fig.~\ref{fig:MMEnhancement_InfRR}(a) where micromotion-enhanced fidelities are observed, and the latter case is an example of where high-fidelity solutions can be found for all micromotion amplitudes. Fig.~\ref{fig:Rep_Rate_Comp}(a) shows that, for gate speeds comparable or faster than a trapping period, accessing the regime of micromotion-enhanced gate fidelities for high micromotion amplitudes (e.g. $q_x=0.3$ and $q_x=0.5$) requires SDK repetition rates on the order of $10^2\omega_0/(2\pi)-10^3\omega_0/(2\pi)$. 
In comparison gates designed for low micromotion environments (e.g. $q_x=0.01$ and $q_x=0.1$) are compatible with lower repetition rates of roughly $10^2\omega_0/(2\pi)$. For a laser repetition rate of $100\omega_0$ and $\Omega_{\rm RF}/(2\pi)=20$Hz, we find that high-fidelity solutions ($F\geq 99.9\%$) across all micromotion environments studied here. This supports the viability of performing fast gates on hyperfine (`clock') qubits, where SDK durations need to be several nanoseconds to suppress off-resonant transitions~\cite{Mizrahi2013c,CW_SDK}.

\section{Error Analysis} 
\label{sec:ErrorAnalysis}

The sensitivity of the fast gate mechanism in the absence of micromotion to experimental errors has been theoretically studied for a range of systematic errors and noise sources in Refs.~\cite{Garcia-Ripoll2003c,Garcia-Ripoll2005c,Zhu2006b,Duan2004a,Bentley2013,Bentley2016a,Taylor2017b,Ratcliffe2018,Gale2020,Mehdi2025,Savill-Brown2025}. Ref.~\cite{Ratcliffe2020} studied the robustness of RF-phase-locked fast gate solutions, however the results cannot be extended to the gate schemes studied in this work where the full temporal structure of the RF cycle is utilized. 

In this section, we study the susceptibility of fast gates to a range of error sources, focusing primarily on pulse area and timing errors in the SDK sequence. For concreteness, we focus on a particular set of high-fidelity gate solutions with a fixed duration of approximately one trapping period, i.e. $t_{\rm g} \approx 2\pi/\omega_0$, tailored for a system with $\Omega_{\rm RF}=40\omega_0$, a SDK repetition rate of $2\pi f_{\rm rep} = 800\omega_0$, and micromotion amplitudes ranging from $q_x=0.01$ to $q_x=0.5$. 

\subsection{Imperfect state-dependent kicks}
\begin{figure} \includegraphics[width=0.8\columnwidth]{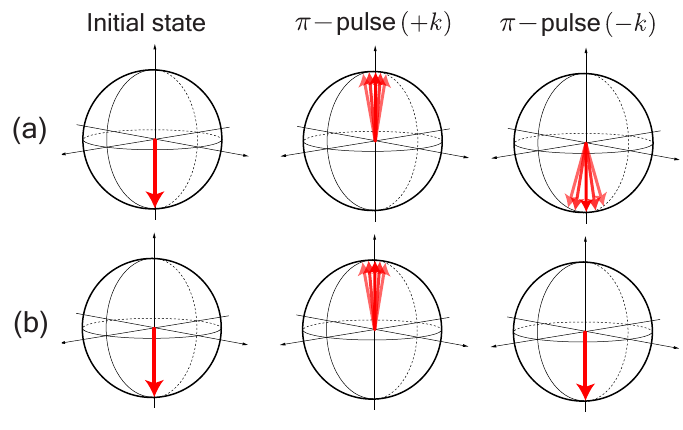}
\caption{Illustration of the effect of SDK pulse area errors on a single qubit. (a) Pulse area errors lead to imperfect population inversion from each $\pi$-pulse, which for an initial spin-state $\ket{\downarrow}$ can be visualised as diffusion of the state at the pole of the single-qubit Bloch sphere. (b) Engineering a $\pi$-phase shift on the counter-propagating $\pi$-pulse -- akin to an `echo' of the Bloch vector which cancels population inversion errors, though motional state errors are not suppressed. \label{fig:SDK_Error_Diagram} }
\end{figure}

Previous work has identified SDK pulse area errors as a key technical limitation to implementing high-fidelity fast gates~\cite{Wong-Campos2017a,Bentley2016a,Gale2020, Savill-Brown2025, Mehdi2025}. These errors can originate from laser power fluctuations, polarization errors, or off-resonant scattering during the SDK, resulting in errant spin-motional states. While the exact nature of SDK errors depends on both the SDK implementation technique, which we discuss in the following section, we can broadly characterize the errors induced by imperfect SDKs as either population transfer errors or motional state errors. 

\begin{figure} \includegraphics[width=1.\columnwidth]{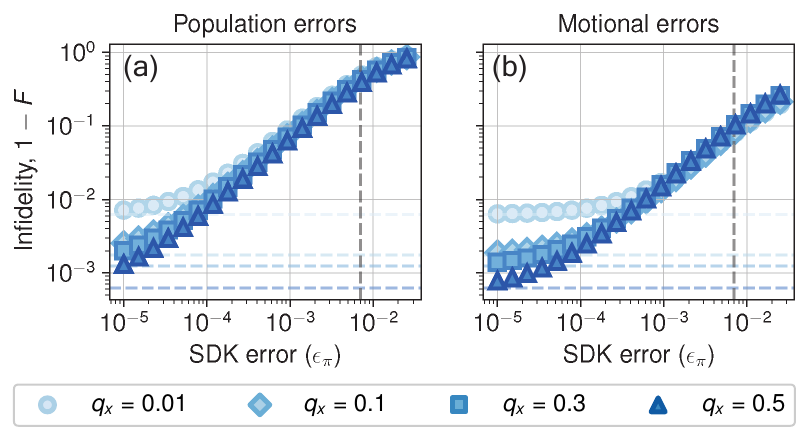}
\caption{The effect of SDK pulse area errors on the fast gate mechanism. (a) Shows the effect of pulse area imperfections in the worst-case scenario where population errors are not suppressed, with the gate fidelity degrading for larger imperfections according to Eq.~\eqref{eq:Fidelity_PopErrors}. (b) Gate infidelity in the case where population errors are suppressed, estimated as the ensemble average over $10^4$ Monte-Carlo samples and a maximum of $m_{\rm max}=4$ errors per SDK sequence.\label{fig:SDK_Errors} }
\end{figure}

Population transfer errors leave the qubit in the wrong internal state, producing a state orthogonal to the target state and leading to two-qubit gate errors that compound with the number of SDKs~\cite{Gale2020}. In particular, for a characteristic SDK error of $\epsilon$, the gate fidelity subject to SDK population transfer imperfections can be bounded as~\cite{Gale2020}:
\begin{align}
\label{eq:Fidelity_PopErrors}
    \mathcal{F}[\epsilon] \geq (1-\mathcal{N}\epsilon)^2\mathcal{F}_0 \,,
\end{align}
where $\mathcal{F}_0$ is the qubit-state-averaged infidelity assuming perfect SDKs, i.e. Eq.~\eqref{eq:StateAvg_Inf}. Figure \ref{fig:SDK_Errors}(b.i) demonstrates that SDK errors of $0.007$~\cite{Johnson2015b} lead to gate fidelities of approximately $70\%$, which is consistent with the order-of-magnitude entangling gate error observed in the experiment of Ref.~\cite{Wong-Campos2017a}.

In principle, population errors can be suppressed by engineering a $\pi$-phase shift within the counter-propagating pulse pair used to implement the SDK (Fig.~\ref{fig:SDK_Error_Diagram}), provided pulse-area errors are strongly correlated between the counter-propagating pulses. This is a reasonable approximation for noise processes (e.g. laser intensity noise or polarization errors) that are slow compared to the short separation between pulse pair constituents, which can be as short as tens of picoseconds. This approach is akin to a spin-echo technique to suppress qubit decoherence, cancelling the spread of the qubit state at the poles of the Bloch sphere as opposed to on the equator, as visualised in Fig.~\ref{fig:SDK_Error_Diagram}. While this method ensures the final internal qubit state is correct, it will not eliminate motional state errors that are correlated with pulse area errors~\cite{Gale2020}. Motional state errors will have some overlap with the ideal final qubit state, which is dependent on the Lamb-Dicke parameter, leading to errors that instead scale as $\mathcal{N}\eta^2$ \cite{Savill-Brown2025} -- suppressing SDK-induced gate errors by up to two orders of magnitude for $\eta\sim 0.1$. A quantitative treatment is required, however, in order to capture the effect of errant motional trajectories beyond this simple estimate. 

We simulate the effects of motional errors on the gate fidelity using a Monte-Carlo technique described in Ref.~\cite{Savill-Brown2025}, where each SDK has a small probability $\epsilon$ of being erroneous (either in the wrong direction or no kick, with equal probability). The mean gate error is then estimated from the weighted sum:
\begin{align}
    \overline{\varepsilon_{\rm G}} = \sum_{m=0}^{m_{\rm max}} \binom{\mathcal{N}}{m} \epsilon^m (1-\epsilon)^{\mathcal{N}-m} \overline{\varepsilon_{\rm G}^{(m)}} \,,
\end{align}
where $\overline{\varepsilon_{\rm G}^{(m)}}$ is the ensemble-averaged infidelity conditioned on exactly $m$ errors per SDK sequence. For $\mathcal{N}\epsilon\ll 1$, we can truncate this sum at a finite order $m_{\rm max}\leq \mathcal{N}$ -- we find $m_{\rm max}=4$ is sufficient for the regimes studied in this work. The results of these simulations are presented in Figure \ref{fig:SDK_Errors}. Fig.~\ref{fig:SDK_Errors}(b) shows the ensemble-averaged gate infidelity for a range of characteristic SDK errors, $\epsilon$. For the current state-of-the-art population transfer error with impulsive SDKs of $0.007$~\cite{Johnson2017b}, we find that average gate fidelities above $90\%$ are achievable, with low-micromotion environments performing best under SDK errors. If SDK errors can be suppressed to the $10^{-3}$ ($10^{-4}$) level, Fig.~\ref{fig:SDK_Errors}(b) suggests gate fidelities above $99\%$ ($99.9\%$) can be reached with operation times of a single radial trapping period. We discuss approaches to improving SDK errors in the following section.

\subsection{Timing and frequency errors}

The impulsive fast gate mechanism relies on carefully timed SDKs that are effectively instantaneous on the timescale of the ion dynamics, and it is thus sensitive to timing noise and frequency noise. This has previously been studied in the context of fast gates addressing the micromotion-free axes of the ion crystal~\cite{Bentley2016a,Ratcliffe2018,Mehdi2025,Savill-Brown2025}, and for RF-phase-locked gate schemes~\cite{Ratcliffe2020}. 

\begin{figure*} \includegraphics[width=2\columnwidth]{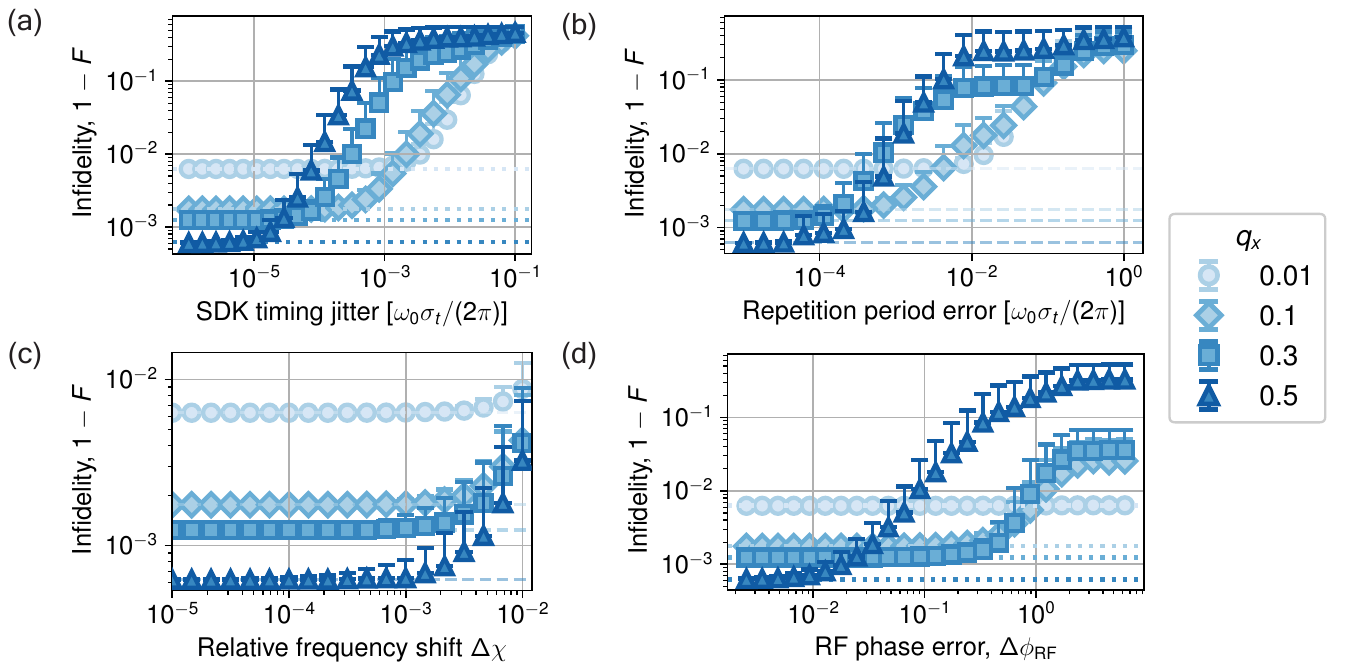}
\caption{Error analysis for select gate solutions with duration of approximately one radial trapping period, i.e. $t_{\rm g} \approx 2\pi/\omega_0$, tailored for a system with $\Omega_{\rm RF}=40\omega_0$, and a SDK repetition rate of $2\pi f_{\rm rep} = 800\omega_0$. In each case we simulate shot-to-shot errors in a key parameter, and report infidelities averaged over $10^3$ noise realisations. Error bars show the range $[\mu,\mu+\varsigma]$ where $\mu$ is the ensemble-averaged mean value and $\varsigma^2$ is variance of the dataset. In each case we find the fidelity of higher-amplitude-micromotion (i.e. larger $|q_x|$) degrades faster with Gaussian noises on (a) SDK timings; (b) the SDK repetition period; (c) the relative frequency splitting between modes; and (d) the initial phase of the RF drive at the time of the first pulse. We find there is no dependence of the $q_x=0.01$ gate solution on the phase of the RF drive, which further justifies its classification as an `effectively secular' gate solution. }
\label{fig:NoiseAnalysis_Timings_StrayFields} 
\end{figure*}

Here we assess the sensitivity of micromotion-enhanced fast gates to a range of timing and frequency errors. The results of this analysis are presented in Figure \ref{fig:NoiseAnalysis_Timings_StrayFields}, and summarised below. In all cases, we find that the fidelity of gates designed for larger micromotion amplitudes, i.e. larger $q_x$ values, exhibits a steeper decline with increasing timing errors.

\subsubsection{SDK timing jitter} We simulate the effect of timing noise on the SDK sequence by sampling random errors on the timings of each pulse from a normal distribution of zero mean and variance $\sigma_t^2$ -- i.e. $t_k  \sim \mathcal{N}\left(t_k^{(0)},\sigma_t^2\right)$ where $\{t_k^{(0)}\}$ are the SDK timings in the absence of noise-- subject to the constraint that $\min{|t_k-t_{k'}|} = 1/f_{\rm rep}$. This effectively sets the SDK repetition rate as a high-frequency cutoff to the noise, which is reasonable given the exceptional frequency stability of modern mode-locked lasers. Figure \ref{fig:NoiseAnalysis_Timings_StrayFields}(a) demonstrates that the gate fidelity degrades faster in higher micromotion environments for increasingly large shot-to-shot timing errors. By comparing the optimal (i.e. no noise) fidelity to the degraded fidelity, we find gate errors at the $10^{-4}$ level are introduced for $\omega_0\sigma_t/(2\pi)\gtrsim 10^{-5}$ in the high micromotion regime, $q_x=0.5$, as compared to $\omega_0\sigma_t/(2\pi)\gtrsim 10^{-4}$ for $q_x=0.1$. For typical radial trapping frequencies of several MHz, this analysis suggests that leveraging high-micromotion environments to improve gate performance requires control pulse timings to be stabilized to the $\mathcal{O}(10)$ps level. Equivalently, the trapping frequency must be characterised to a precision of better than one part in ten thousand.

\subsubsection{SDK repetition period error} Timing errors can arise due to shot-to-shot variations of the SDK repetition rate. Assuming the gate design is correctly calibrated to the mean SDK repetition period, we can model the effect of this noise by sampling the SDK repetition period $t_{\rm rep}$ from a normal distribution $ \mathcal{N}\left(t_{\rm rep}^{(0)},\sigma_t^2\right)$, where $1/t_{\rm rep}^{(0)}$ is the SDK repetition rate assumed in the gate design. This noise only affects the temporal separation of SDKs within each group while keeping the time delay between the first pulse of consecutive SDK groups fixed.  Figure \ref{fig:NoiseAnalysis_Timings_StrayFields}(b) illustrates that the advantage of high micromotion environments is lost for repetition period errors larger than $\omega_0\sigma_t/(2\pi)\approx 10^{-3}$, where the gate infidelity approaches the $1\%$ level. In order to suppress SDK-repetition-period induced errors below $10^{-4}$, Fig.~\ref{fig:NoiseAnalysis_Timings_StrayFields}(b) shows that SDK repetition period jitter must be roughly $10$ps or smaller, consistent with the estimate for SDK timing jitter studied above.

\subsubsection{Stray fields} Next we consider the effect of stray electric fields which alter the motional mode structure of the ion crystal. Provided the stray fields vary slowly with respect to the RF drive, their effect on the fidelity of fast gate solutions can be modelled as a shot-to-shot variations in the relative frequency splitting between the secular modes, $\chi$~\cite{Ratcliffe2020,Mehdi2025}. For simplicity we assume the noise is Gaussian distributed, i.e. $\chi\sim \mathcal{N}\left(-0.014,(\Delta \chi)^2\right)$; ensemble averaged gate errors are presented in Fig.~\ref{fig:NoiseAnalysis_Timings_StrayFields}(c) as a function of $\Delta \chi$. Unlike the timing errors studied above, gates designed in low and high micromotion environments degrade at similar rates, with gate errors of order $10^{-4}$ ($10^{-2}$) induced for relative frequency shifts of order $0.01\%$ ($1\%$). For typical MHz trapping frequencies, this corresponds to stabilization of the RF pseudopotential to within several hundred Hz.

\subsubsection{RF phase error} Last, we consider the effect of miscalibration in the phase of the RF drive, which needs to be well defined at the time of the first pulse for the fast gate protocols presented in this work. We model this effect as shot-to-shot variations in the RF drive phase sampled from a normal distribution with zero mean and variance $\Delta\phi_{\rm RF}^2$, with ensemble-averaged gate errors presented in Fig.~\ref{fig:NoiseAnalysis_Timings_StrayFields}(d). There is no degradation of gate fidelity in the case of the lowest micromotion amplitude, $q_x=0.01$, for phase errors spanning the full $2\pi$ range. This is consistent with the interpretation of this regime as `effectively secular'. For micromotion amplitudes $q_x>0.1$ the gate fidelity degrades faster in higher micromotion environments. In order to maintain gate errors below one part per thousand, the RF phase must be known and stabilized to below $5\%$ for $q_x=0.5$ and below $10\%$ for $q_x=0.1-0.3$.

\subsection{Motional heating}
In addition to motional dephasing due to stray fields, heating of the motional modes -- e.g. from noise on electrode surfaces~\cite{Turchette2000} -- can degrade gate fidelity. Previous studies have shown that fast gates are naturally robust to motional-mode heating due to their high speeds compared to typical heating rates~\cite{Taylor2017b,Savill-Brown2025}. Concretely, we can estimate the heating-induced-error to be roughly $t_{\rm g}\dot{\bar{n}}\ll 1$, where $\dot{\bar{n}}$ is the largest heating rate across all excited modes. For microsecond gate durations, the induced error is less than $10^{-4}$ for heating rates of order $\dot{\bar{n}}\sim 100$~quanta/s typical of surface-electrode traps~\cite{Bruzewicz2019}.

\section{Experimental considerations}

In this section, we discuss experimental considerations for the implementation of the SDK-based fast gate mechanism, noting there has only been a single experimental demonstration with $76\%$ fidelity~\cite{Wong-Campos2017a}. SDK errors are expected to be the key limiting factor to high-fidelity fast gates, as discussed in the previous section. A judicious choice of SDK implementation scheme is therefore essential for improving the fast gate fidelity.

\subsection{SDK implementation} An advantage of SDK-based fast gates is that the entangling gate scheme is qubit-species-agnostic beyond the ability to perform high-fidelity SDKs on each addressed ion~\cite{Mehdi2025}. In addition, the large timescale separation between the motional dynamics of the ion crystal (microseconds) and the duration of a single SDK (picoseconds to nanoseconds) means that a broad range of SDK implementation schemes can be considered. Table~\ref{tab:SDK_Implementation} summarises experimentally demonstrated SDK schemes, with state-of-the-art errors on the order of $1\%$.

For implementing high-fidelity SDKs on hyperfine `clock' qubits, nanosecond sequences of $\lesssim 10$ picosecond pulses can be numerically optimized to suppress motional diffraction into unwanted momentum orders~\cite{Mizrahi2014,Wong-Campos2017a,Savill-Brown2025}. The extended duration of this SDK scheme, as compared to single pulse schemes for Zeeman qubits~\cite{Madsen2006,Campbell2010b,Putnam2024}, constrains the effective SDK repetition rate to be an order of magnitude lower than the repetition rate of the pulsed laser source. The analysis in Section~\ref{sec:RepRate} suggests that high-fidelity gate solutions can still be found when constraining the SDK repetition rate to be $\mathcal{O}(100)\omega_0$, with durations as short as $2.5$ trapping periods (e.g. $500$ns for $\omega_0=2\pi\times 5$MHz). The design and implementation of high-fidelity SDKs with durations comparable to the period of the RF drive should be a focus of future theoretical and experimental work.

We propose implementing radial SDKs on barium ($\text{Ba}^+$) ions utilizing the ground-state Zeeman qubit. The experimental configuration centers on a pulsed laser source, split into a pair of counter-propagating beams. With the combination of a pulsed laser source, electro-optic modulators (EOMs) and carefully selected polarisation optics one can arbitrarily shape sequences of pulses and switch pulse directionality to provide forward and backward SDKs. The advantage of using Zeeman qubits over hyperfine encodings is the ability to implement an SDK with a single picosecond pulse, which enables SDK delivery at the repetition rate of the pulsed laser source which can be as large as several GHz~\cite{Heinrich2019b}. For typical radial trapping frequencies of several MHz, this would enable high-speed gates in the sub-trap-period regime which require SDK repetition rates ranging from $500-1000\times$ the secular trapping frequency. Concretely, for a radial trapping frequency of $5$MHz with $q_x=0.5$ and a $5$GHz laser repetition rate~\cite{Heinrich2019b}, Figure~\ref{fig:Rep_Rate_Comp} demonstrates that gate durations below $200$ns (a single trapping period) are achievable with a fidelity of $99.99\%$.

\begin{table*}[]
\begin{tabular}{@{}lllll@{}}
\toprule
\multicolumn{1}{c}{Implementation Scheme} & \multicolumn{1}{c}{Duration} & \multicolumn{1}{c}{Primary limitation}          & \multicolumn{1}{c}{Qubit type} & \multicolumn{1}{c}{Best fidelity demonstrated}                                                                                            \\ \midrule
Single resonant photons                   & $\mathcal{O}(10)$ps          & Spontaneous scattering & O,H,Z     & 0.94~\cite{Shimizu2021}, 0.96~\cite{Heinrich2019b}, 0.99~\cite{Guo2022e} \\
Two-photon (single pulse)                     & $\mathcal{O}(10)$ps          & Differential light shifts               & Z                         & 0.95~\cite{Putnam2024}                                                                                              \\
Two-photon (pulse train)               & $\mathcal{O}(1)$ns           & Multi-photon motional diffraction       & H                      & 0.991~\cite{Johnson2017b}$^\dagger$                                    \\ \bottomrule
\end{tabular}\label{tab:SDK_Implementation}\caption{Summary of SDK implementation schemes demonstrated in trapped-ion systems. Qubit types: optical (O), hyperfine (H), Zeeman (Z). Schemes using a single resonant photon are limited by spontaneous scattering from the excited state, whereas two-photon driving enables high-fidelity transfer between split ground-state levels (hyperfine or Zeeman) with suppressed scattering rates. ${}^\dagger$Ref.~\cite{Johnson2017b} reports a fidelity of spin population transfer of $0.993$, which does not include motional diffraction errors.  }
\end{table*}

\subsection{Suppressing SDK errors} In order for fast gates to be implemented with suitably high fidelities for quantum computation, future work must focus on reducing the error of SDKs below the current demonstrations (roughly $1\%$). Simulations have shown that controlling the timing of ultrafast pulses enables SDK fidelities of $99.99\%$~\cite{Mizrahi2014,Putnam2024}, and similar optimal control models strongly suggest that this can be further improved with amplitude and phase control. These controls should enable designs that are robust to polarization and intensity noise. In a similar vein, multiple SDKs could be concatenated to form a composite-pulse SDK that is first- or second-order insensitive to particular error sources~\cite{Wu2023}. Such an approach has been developed for trapped-ion systems, though has yet to be extended to ultrafast spin-motional control. It also may be possible to use dark-state adiabatic protocols such as STIRAP~\cite{Vitanov2017}, which asymptotically achieve ideal population transfer provided the adiabatic condition is met. 

One of the challenges of ultrafast spin-motional control is the inherent broadband nature of pulses from mode-locked lasers, which leads to difficulty with differential light-shifts due to undesired off-resonant frequency components. Furthermore, producing non-uniform pulse timings requires optical elements such as nested delay lines and dispersive elements. In contrast, continuous-wave (CW) lasers are a viable alternative that could be implemented using an EOM to achieve sub-nanosecond shaping of both pulse amplitude and phase. This simultaneously enables amplitude and phase-modulated pulses, needed for high-fidelity atomic transitions, and programmable non-uniform timings. Liu \emph{et al.}~recently proposed a Raman SDK scheme using amplitude-modulated CW pulses. They predicted infidelities below $10^{-5}$ in the presence of micromotion, with peak Rabi frequencies ($\sim 100$MHz) more than fifty times lower than equivalent pulsed laser implementations~\cite{CW_SDK}.

\subsection{Choice of qubit species} 
Historically, species such as Ytterbium ($\text{Yb}^+$) and Calcium ($\text{Ca}^+$) have been dominant in trapped-ion quantum computing, but recent interest has shifted toward Barium ($\text{Ba}^+$) ions due to several key advantages~\cite{Dietrich2009}. These benefits include low spontaneous emission~\cite{Moore2023}, high state preparation and measurement (SPAM) fidelity~\cite{Sotirova2024}, the ability to use multiple qubit encodings for the Optical Metastable Ground state (OMG) architecture~\cite{Allcock2021}, desirable visible wavelengths for integrated optics that can be converted into telecommunication bands~\cite{Siverns2017}, and compatibility with quantum networking protocols~\cite{Mehdi2025,OReilly2024,Drmota2023a,Main2025}. 

Crucially for the design of fast gates, $\text{Ba}^+$ possesses a favorable stimulated Raman transition wavelength at 532 nm, unlike $\text{Yb}^+$ and $\text{Ca}^+$, which typically require lasers in the ultraviolet (UV) regime. This is particularly important as polarization instabilities associated with high-power UV pulses passing through dispersive optical elements is quoted as the leading limitation to achieving high-fidelity entanglement in the experiment of Ref.~\cite{Wong-Campos2017a}. Furthermore, there are practical advantages to the use of 532 nm wavelength driving fields because of the wide commercial availability of high-power lasers -- both pulsed and continuous wave (CW) -- at this wavelength.

\subsection{Optical access and ion addressing} Our work focuses on radial gates since one of the main experimental constraints when implementing laser-based gate operations on trapped ion systems is optical access. 
In both macroscopic linear traps and surface electrode traps, radial ion addressing is often preferred or imposed by the geometry of the trap~\cite{Bruzewicz2019}.
Based on Eq.~\eqref{eq:StateAvg_Inf}, the gates are first order insensitive to beam misalignment. 

While individual addressing is not required for fast gates in a two-ion crystal, using a single global beam requires more laser power in order to deliver the same intensity to each ion. Furthermore, individual addressing of each ion provides further advantages for mixed-species crystals where each ion qubit can be addressed with a pulse tailored to the specific atomic transitions of each species or isotope~\cite{Mehdi2025}. 

\subsection{Trapping geometry}
In this work, we chose trapping parameters $(a_\alpha, q_x)$ that maintain the same secular frequency across all micromotion environments, enabling a controlled comparison between different $q_x$ values [see Appendix~\ref{app:opti-params} and Fig.~\ref{fig:Schematic}(c)]. For the highest micromotion amplitudes studied ($q_x = 0.5$), the parameter $a_\alpha \approx -0.1$ requires DC voltages of order~\cite{Leibfried2003b} $U_{\rm DC} \sim |a| m \Omega_{\rm RF}^2 r_0^2 / (4e) \approx 22$~V ($87$~V) for an ion-electrode distance of $r_0 = 100\,\mu$m ($200\,\mu$m) assuming $^{133}\text{Ba}^+$ ions in a trap with $\Omega_{\rm RF} = 2\pi \times 40$~MHz. While achievable, these voltages are at the upper end of what is typical in current experiments. We note, however, that the large $|a_\alpha|$ values are a consequence of our comparison methodology, not a requirement for micromotion enhancement: high $q_x$ values can equally be achieved with $|a| \approx 0$ at the cost of higher secular frequencies, which are accessible in miniaturised 3D-printed ion traps~\cite{Xu2025}.

\section{Conclusion and Outlook}

In this manuscript, we have presented a detailed study on fast two-qubit gates using SDKs implemented on the micromotion-sensitive radial modes of a two-ion crystal. We have demonstrated that micromotion can be effectively incorporated into a quantum control framework, which enables SDKs sequences that realise high-fidelity gate solutions to be identified. Furthermore, we have demonstrated the existence of experimentally-feasible parameter regimes where larger micromotion amplitudes improve both the speed and fidelity of fast gate solutions; the latter by up to two orders of magnitude. Through detailed analysis of numerically-optimized gate solutions, we studied the susceptibility of radial fast gates to various sources of error, finding that gates designed with larger micromotion amplitudes are more sensitive to timing and frequency errors. While most error sources can be suppressed in current experiments, compounding SDK errors limit the achievable gate fidelity above $0.9$ with experimentally demonstrated SDK fidelities. With further advances to ultrafast spin-motional control~\cite{CW_SDK}, this could be pushed into the $>0.999$ regime relevant for quantum information processing while maintaining quantum logic rates of $1-10$MHz, without requiring cooling of the ion crystal to any particular regime (e.g. the Lamb-Dicke regime for M\o{}lmer-S\o{}renson-type gates~\cite{Sorensen2000}).

\subsection{Future work}

There are a number of avenues for future theoretical investigations of the fast gate mechanism. The first is to investigate limits to the theoretical gate fidelity, i.e. in the absence of control errors. This could be done systematically by adding additional complexity to the SDK pulse sequences, e.g. by introducing multi-loop schemes that suppress residual spin-motional entanglement errors as has been demonstrated in the context of adiabatic control of trapped-ion qubits~\cite{Hayes2012,Bentley2020}. Further extensions may focus on identifying gate schemes (i.e. particular SDK sequences) that maintain fidelity in the presence of shot-to-shot variations of experimental parameters, e.g. within the framework of robust quantum control~\cite{Milne2020,Vedaie2023}.

High-speed and high-fidelity radial entangling gates provide new possibilities for scaling up trapped-ion quantum computers. One example is architectures based on long ion chains~\cite{Debnath2016,Figgatt2019,Grzesiak2020,Chen2024}, where radial modes have several advantages for entangling operations. First, the radial confinement remains strong (typically $1-10$MHz) as the chain is scaled up, leading to lower phonon occupations of the radial modes of long chains~\cite{Zhu2006b}. Third, shuttling of ion crystals is typically done axially which leads to heating of the axial modes~\cite{Palmero2014,Pino2021,Sterk2022,Burton2023}; therefore using radial modes for logic gates reduces the need for frequent cooling.

Fast gates also present new possibilities for shuttling-free trapped-ion architectures~\cite{Ratcliffe2018,Mehdi2020b}, as entangling gates can be performed between ions in neighbouring traps using large-momentum transfer from groups of concatenated SDKs~\cite{Gale2020}. While more experimentally challenging, such architectures -- which naturally extend to two-dimensional arrays~\cite{Mehdi2020b,Wu2020b} -- provide a pathway for overcoming the key latency bottlenecks of ion shuttling: Doppler and motional ground state re-cooling ~\cite{Pino2021,Moses2023d}. Micromotion must explicitly be taken into account in the design of quantum logic operations in these architectures, as micromotion can only be eliminated on a single axis of the trap~\cite{Leibfried2003b}. Indeed, some microtrap arrays employ cylindrical trapping configurations in which there is no micromotion-free axis~\cite{Kumph2016}.

Moving beyond linear trapping configurations presents further opportunities for trapped-ion quantum computation. For example, two-dimensional ion crystals -- which can host hundreds of ions~\cite{Szymanski2012} -- have emerged as a candidate for scaling up the number of ions that can be confined together while retaining the ability for individually addressed quantum logic operations between any two ions~\cite{Hou2024}. Micromotion becomes more significant in two-dimensional and three-dimensional ion crystals, with ions further from the center of the crystal experiencing larger amplitude modulation~\cite{Wu2021}. The results of this work -- larger amplitude micromotion can improve both gate speed and fidelity -- motivate the study of fast gates in two- and three-dimensional ion crystals, which present interesting opportunities for scaling up trapped-ion processors to thousands of physical qubits.

\section*{Acknowledgments} 
This research was undertaken with the assistance of supercomputing resources and services from the National Computational Infrastructure, which is supported by the Australian Government.

\appendix
\section{Quantum dynamics in radio-frequency traps \label{sec:BackgroundTheory}} 
In this appendix we briefly review the theory of ion motion in radio-frequency traps, closely following the derivation in Ref.~\cite{Leibfried2003b}. We focus on the linear Paul trap configuration: static trapping along the linear axis of the trap ($z$) and dynamic trapping in the $x-y$ plane. We further assume excitation of the ion crystal purely along the $x$ axis, for which the RF potential can be expressed as
\begin{align}
\notag	V_{\rm RF}(x,t) = \frac{1}{2}m(\Omega_{\rm RF}/2)^2\left[a_x-2q_x\cos(\Omega_{\rm RF} t +\phi_{\rm RF}) \right]x^2 \,,
\end{align} 
where $m$ is the mass of the ion, $\Omega_{\rm RF}$ is the frequency of the RF drive with phase $\phi_{\rm RF}$ at $t=0$. The parameters $a_x$ and $q_x$ are trapping parameters proportional to the strength of the DC and AC trapping fields, respectively~\cite{Leibfried2003b}.

Then, assuming two ions fixed to their equilibrium positions in the $y$ and $z$ axes, -- $y^{(1)}_0=y^{(2)}_0=0$ and $z^{(1)}_0=-z^{(2)}_0=d/2$, the potential energy of the two-ion system along the radial ($x$) axis is given by:
\begin{align}
	V(x^{(1)},x^{(2)},t) = &\sum_{j=1,2}V_{\rm RF}(x^{(j)},t) \\ \notag &+ \frac{e^2}{4\pi\epsilon_0}\frac{1}{\sqrt{(x^{(1)}-x^{(2)})^2+d^2}},
\end{align}
where $e$ is the elementary charge. This potential has a stationary point at $x^{(1)}=x^{(2)} = 0$, which corresponds to ion equilibria along the null of the RF trap (see Fig.~\ref{fig:Schematic}). 
Assuming small excursions in phase space, such that we can expand the Coulomb term to quadratic order in $(x^{(1)}-x^{(2)})/d$, and transforming into centre-of-mass (CM) and relative/stretch (SR) coordinates, i.e. $	x_{\rm CM/SR} = (x^{(1)}\pm x^{(2)})/\sqrt{2}$, the potential energy of the system may be expressed as:
\begin{align}
		V_{\rm RF}(x_{\rm CM},x_{\rm SR})&\approx \frac{1}{2}m \sum_{\alpha={\rm CM,SR}}\lambda_\alpha(t) x_\alpha^2
\end{align}
where $\lambda_\alpha(t) \equiv (\Omega_{\rm RF}/2)^2[a_{\alpha} - 2 q_x\cos(\Omega_{\rm RF} t+\phi_{\rm RF})]$ in terms of the mode-dependent trapping parameter $a_\alpha = a_x-4\Gamma_\alpha(d)/(m \Omega_{\rm{RF}}^2)$. 
For a one-dimensional system of $N=2$ ions in a common harmonic potential, $\Gamma_{\rm CM}(d)=0$ and
\begin{align}
	\Gamma_{\rm SR} = \frac{e^2}{4\pi\epsilon_0 d^3} \,.
\end{align}

After linearization of the Coulomb potential around the equilibrium configuration of the two-ion crystal, the Hamiltonian describing the motional degrees of freedom can be expressed as a sum over each mode -- i.e. $\hat{H}_{\rm mot} = \sum_\alpha \hat{H}_{{\rm mot},\alpha}$ where the index $\alpha$ labels the motional modes and:
\begin{align}
\label{eq:Hmot_alpha}
    \hat{H}_{{\rm mot},\alpha}(t) = \frac{\hat{p}_\alpha^2}{2m} + \frac{1}{2}m\lambda_\alpha(t)\hat{x}_\alpha^2 \,.
\end{align}
Here we adopt a Heisenberg picture approach~\cite{Leibfried2003b, Ji1995, Wu2021}, where the position and momentum operators of each motional mode evolve according to the equations of motion:
\begin{align}
       \frac{d\hat{x}_\alpha}{dt} &= \frac{\hat{p}_\alpha}{m} \,,\quad
       \frac{d\hat{p}_\alpha}{dt}= -m\lambda_\alpha(t)\hat{x}_\alpha\,.
\end{align} 
These equations of motion admit the following solution~\cite{Leibfried2003b}:
\begin{subequations}
\label{eq:HeisenbergPicture_XandP}
\begin{align}
	\hat{x}_\alpha(t) &= \sqrt{\frac{\hbar}{2m\omega_\alpha}}  \left(\hat{a}_\alpha u^*_\alpha(t) + \hat{a}^\dag_\alpha u_\alpha(t)\right) \,, \\
	\hat{p}_\alpha(t) &= \sqrt{\frac{\hbar m}{2\omega_\alpha}}\left(\hat{a}_\alpha \dot{u}^*_\alpha(t) + \hat{a}^\dag_\alpha \dot{u}_\alpha(t)\right) \,,
\end{align}
\end{subequations}
in terms of the time-dependent mode function:
\begin{align}
\label{eq:MM_modefunctions}
	u_\alpha(t) = \sum_{n=-\infty}^\infty C_{\alpha,n} e^{i\beta_\alpha\Omega_{\rm RF} t/2}e^{in(\Omega_{\rm RF}t + \phi_{\rm RF})} \,,
\end{align} 
which is a solution to the Mathieu-Hill equation, $\ddot{u}_\alpha(t) = - \lambda_\alpha(t) u_\alpha(t) $. These mode functions include a low frequency component, which we refer to as the \textit{secular} mode frequency $\omega_\alpha \equiv \beta_\alpha\Omega_{\rm RF}/2$, 
and high-frequency components at integer multiples of the RF drive -- i.e. micromotion. 
The Fourier coefficients in Equation~\eqref{eq:MM_modefunctions} satisfy the recurrence relation
\begin{align}
\label{eq:RecursionRelation_Mathieu_Cn}
    C_{\alpha,n+1}-D_{\alpha,n} C_{\alpha,n} +C_{\alpha,n-1} = 0 \,,
\end{align}
where $D_{\alpha,n} \equiv [a_\alpha-(2n+\beta_\alpha)^2]/q_x$ in terms of the (mode-dependent) Floquet exponent $\beta_\alpha$. Eq.~\eqref{eq:RecursionRelation_Mathieu_Cn} can be re-arranged to obtain continued fraction solutions for the coefficients:
	\begin{align}
\label{eq:CFraction_Cns_Descending}
    C_{\alpha,n} &= \frac{C_{\alpha,n+1}}{D_{\alpha,n} - \frac{1}{D_{\alpha,n-1}-\frac{1}{D_{\alpha,n-2}-\dots}}} \\ &= \frac{C_{\alpha,n-1}}{D_{\alpha,n} - \frac{1}{D_{\alpha,n+1}-\frac{1}{D_{\alpha,n+2}-\dots}}} \notag \,.
\end{align}
We take $C_{\alpha,0}=1$ without loss of generality, as its value can be absorbed into the initial conditions for the Heisenberg operators $\{\hat{x}_\alpha(t_0),\hat{p}_\alpha(t_0)\}$. The above continued fraction solutions can be re-arranged to obtain a self-consistency equation for the characteristic exponent $\beta_{\alpha}$:
\begin{align}
 \beta_\alpha^2 &=a_\alpha-q_x\left(\frac{1}{D_{\alpha,1} - \frac{1}{D_{\alpha,2}-\dots}}+ \frac{1}{D_{\alpha,-1}-\frac{1}{D_{\alpha,-2}-\dots}}\right) \,,
\end{align}
which can be solved numerically, e.g. using the \textsc{MathieuCharacteristicExponent} function in Mathematica.

\section{Derivation of fast gate condition equations in the presence of micromotion }\label{app:FGDerivation}

In this appendix we present a detailed derivation of the fast gate condition equations presented in the main text, Equations~\eqref{eq:Phase_ConditionEquations}--\eqref{eq:Y_ConditionEquations}. 

\subsection{ Dimensionless motional mode quadratures}
We begin by expressing the motional Hamiltonian in non-dimensional form. We choose to non-dimensionalize the position and momentum operators for each motional mode -- given in Equation \eqref{eq:HeisenbergPicture_XandP} -- in terms of the harmonic oscillator lengthscale $l_{0,\alpha}^2=\hbar/(m\omega_\alpha)$, i.e.
\begin{subequations}
\label{eq:app:HeisenbergXY_Dimensionless}
\begin{align}
	\hat{X}_\alpha(t) \equiv\hat{x}_\alpha(t)/l_{0,\alpha} &= \frac{\hat{a}_\alpha u^*_\alpha(t) + \hat{a}^\dag_\alpha u_\alpha(t)}{\sqrt{2}}  \,, \\
	\hat{Y}_\alpha(t) \equiv l_{0,\alpha}\hat{p}_\alpha(t)/\hbar &= \frac{\hat{a}_\alpha \dot{u}^*_\alpha(t) + \hat{a}^\dag_\alpha \dot{u}_\alpha(t)}{\sqrt{2}\omega_\alpha} \,.
\end{align}
\end{subequations}
We refer to these operators as the (dimensionless) position and momentum quadrature of the $\alpha$-th mode, respectively, as they satisfy the canonical commutation relation $[\hat{X}_\alpha,\hat{Y}_{\alpha'}]=i\delta_{\alpha,\alpha'}$.

The motional Hamiltonian of the system, Eq.~\eqref{eq:Hmot_alpha}, can then be written in the form:
\begin{align}
    \hat{H}_{\rm mot}(t) = \sum_\alpha \frac{\hbar \omega_\alpha}{2}\left(\hat{Y}_\alpha^2 + \frac{\lambda_\alpha(t)}{\omega_\alpha^2}\hat{X}^2\right) \,,
\end{align}
from which we obtain the Heisenberg equations of motion (EOMs):
\begin{subequations}
    \begin{align}
        \dot{\hat{X}}_\alpha &= \omega_\alpha \hat{Y}_\alpha \,, \\ 
\label{eq:app:dimensionless_eom_Y}        \dot{\hat{Y}}_\alpha &= -\frac{\lambda_\alpha(t)}{\omega_\alpha }\hat{X}_\alpha \,.
    \end{align}
\end{subequations}
Combining these two equations gives the (quantized) Mathieu-Hill equation, $\ddot{\hat{X}}_\alpha(t)+\lambda_\alpha(t)\hat{X}_\alpha(t) = 0$, the solution of which is given by Eq.~\eqref{eq:app:HeisenbergXY_Dimensionless} in terms of the Mathieu-Hill mode function Eq.~\eqref{eq:MM_modefunctions}.
    
\subsection{ Semiclassical motion due to state-dependent kicks}
In this section, we derive an analytic expression for the state-dependent motion of the ion crystal. Our analysis will be performed in the semiclassical limit, where the Heisenberg-picture operators in Eq.~\eqref{eq:HeisenbergPicture_XandP} are replaced with their expectation value with respect to an initial motional state, i.e. 
\begin{align}
    \hat{X}_\alpha(t) \rightarrow X_\alpha (t)&\equiv \langle \xi_\alpha| \hat{X}_\alpha |\xi_\alpha\rangle = \frac{\xi_\alpha u^*_\alpha(t) + \xi_\alpha^* u_\alpha(t)}{\sqrt{2}} \,,
\end{align}
where $\ket{\xi_\alpha}$ is a coherent state of the $\alpha$-th mode (i.e. $\hat{a}_\alpha\ket{\xi_\alpha}=\xi_\alpha\ket{\xi_\alpha}$). The amplitude of the coherent state, $\xi_\alpha$, encodes the initial conditions of the motional evolution.

To correctly describe the semiclassical ion trajectories associated with each 2Q basis state, the evolution must be treated piecewise: each SDK enforces a new boundary condition for the following period of free evolution. This requires us to express the semi-classical evolution of each mode in terms of arbitrary initial conditions $X_\alpha(t_0)$ and $Y_\alpha(t_0)$
\begin{subequations}
    \begin{align}
	X_\alpha(t_0) &= \frac{\xi_\alpha+\xi_\alpha^*}{\sqrt{2}}u_{\alpha,C}(t_0) + i\frac{\xi_\alpha^*-\xi_\alpha}{\sqrt{2}} u_{\alpha,S}(t_0)  \,, \\
	\omega_\alpha Y_\alpha(t_0) &= \frac{\xi_\alpha+\xi_\alpha^*}{\sqrt{2}} \dot{u}_{\alpha,C}(t_0) + i\frac{\xi_\alpha^*-\xi_\alpha}{\sqrt{2}} \dot{u}_{\alpha,S}(t_0) \,.
\end{align}
\end{subequations}
where we have decomposed the mode function, Eq.~\eqref{eq:MM_modefunctions}, into real (cosine) and imaginary (sinusoidal) components: $u_\alpha(t) = u_{\alpha,C}(t)+iu_{\alpha,S}(t)$. Re-arranging the above gives the following expression for the coherent state amplitude in terms of the initial conditions $\{X_\alpha(t_0),Y_\alpha(t_0)\}$:
\begin{widetext}
    \begin{align}
    \xi_\alpha =  \frac{1}{\sqrt{2}}\frac{\left(\omega_\alpha Y_\alpha(t_0) u_{\alpha,S}(t_0)-X_\alpha(t_0) \dot{u}_{\alpha,S}(t_0) \right) + i\left(X_\alpha(t_0) \dot{u}_{\alpha,C}(t_0)-\omega_\alpha(t_0) Y_\alpha(t_0) u_{\alpha,C}(t_0)\right)}{u_{\alpha,S}(t_0)\dot{u}_{\alpha,C}(t_0)-u_{\alpha,C}(t_0)\dot{u}_{\alpha,S}(t_0)} \,.
\end{align}
\end{widetext}
Due to the linearity of the dynamics the evolution for the full gate is given by the sum of solutions that satisfy the boundary conditions of each SDK. 
The motional dynamics induced by a single SDK at time $t_j$ can be described by the initial conditions $X_\alpha(t_j) = 0$ and $ Y_\alpha(t_j) = 2^{3/2}z_j \eta_\alpha (b_\alpha^{(A)}\hat{\sigma}_z^{(A)}+b_\alpha^{(B)}\hat{\sigma}_z^{(B)})$~\footnote{The result is independent of the particular choice of initial condition due to the linearity of the equations of motion; thus the initial condition $X_\alpha = 0$ can be assumed without loss of generality.}. The latter is consistent with the transformation of the coherent state amplitude from a single SDK with wavevector $2k$, i.e. $\beta \rightarrow \beta + 2iz_j\hat{\eta}_\alpha^{(A,B)}$, where we have introduced the notation $\hat{\eta}_\alpha^{(A,B)} = \eta_\alpha (b_\alpha^{(A)}\hat{\sigma}_z^{(A)}+b_\alpha^{(B)}\hat{\sigma}_z^{(B)})$ for brevity. For a series of $\mathcal{N}$ SDKs at times $\{t_1,t_2,\dots,t_\mathcal{N}\}$ with directions $z_j = \pm 1$ for $j=1,\dots,\mathcal{N}$, the general solution for the position at some time $t>t_1$ is:
\begin{align}
\notag	X_\alpha (t)=  \sum_{j=1}^\mathcal{N} & \frac{H(t_j-t)  z_j \sqrt{2}\hat{\eta}_\alpha^{(A,B)}}{\rho_\alpha(t_j)}\big[u_{\alpha,S}(t-t_j)u_{\alpha,S}(t_j)\\  &\;\;\;\;\;\;\;\;\;\;\;\;\;\;\;\;\;-u_{\alpha,C}(t-t_j)u_{\alpha,C}(t_j) \big]
\end{align}
where $H(t)$ is the Heaviside step function and we have defined $\rho_\alpha(t_j) = [u_c(t_j)u_s'(t_j)-u_s(t_j)u_c'(t_j)]/\omega_\alpha$.

To compare the motional dynamics to the case of purely harmonic evolution (i.e. no micromotion), it is convenient to separate out the secular evolution at frequency $\omega_\alpha=\beta_\alpha\Omega_{\rm RF}/2$ in Eq.~\eqref{eq:MM_modefunctions} from high-frequency components at integer multiples of $\Omega_{\rm RF}$. That is, we define the following cosine-like and sine-like functions:
\begin{align}
	f_{\alpha,S}(t) &\equiv \sum_{n}C_n\sin(n\Omega_{\rm RF} t+n\phi) \,, \\
		f_{\alpha,C}(t) &\equiv \sum_{n}C_n\cos(n\Omega_{\rm RF} t+n\phi) 
\end{align}
in terms of which we can write the real and imaginary components of $u_\alpha(t)$ as:
\begin{align}
	u_{\alpha,S}(t) &= \sin(\omega_\alpha t) f_{\alpha,C}(t)+\cos(\omega_\alpha t) f_{\alpha,S}(t) \,, \\
	u_{\alpha,C}(t) &= \cos(\omega_\alpha t) f_{\alpha,C}(t)-\sin(\omega_\alpha t) f_{\alpha,S}(t) \,.
\end{align}
Substituting this into the result above gives  $\rho_\alpha(t_j) = 1+[f_{\alpha,C}(t_j)\dot{f}_{\alpha,S}(t_j)-f_{\alpha,S}(t_j)\dot{f}_{\alpha,C}(t_j)]/\omega_\alpha$ and 
\begin{align}
\notag	X_\alpha (t)=&  \sum_{j=1}^\mathcal{N} \frac{H(t_j-t)  z_j \sqrt{2}\hat{\eta}_\alpha^{(A,B)}}{\rho_j}\big[\cos(\omega_\alpha [t-t_j]) \\ &\times \mu^{(s)}_\alpha(t,t_j)+ \sin(\omega_\alpha [t-t_j])\mu^{(c)}_\alpha(t,t_j) \big] \,,
    \label{eq:appendix:Xalpha_solution}
\end{align}
in terms of the symmetric (cosine-like) and anti-symmetric (sine-like) variables $\mu^{(c)}(t,t')$ and $\mu^{(s)}(t,t')$ defined in Equation \eqref{eq:TensorsDef_mu} of the main text. The analytic solution for the momentum quadrature of each mode is simply given in terms of the time derivative of the above solution by Eq.~\eqref{eq:app:dimensionless_eom_Y}, i.e. $Y_\alpha (t) = \dot{X}_\alpha(t)/\omega_\alpha$. To obtain Eq.~\eqref{eq:ConditionEquations} of the main text, we evaluate the mode quadratures at the end of the gate $t=t_{\rm g}$ and separate out a factor of $(b_\alpha^{(A)}\hat{\sigma}_z^{(A)}+b_\alpha^{(B)}\hat{\sigma}_z^{(B)})$ from $\hat{\eta}_\alpha^{(A,B)}$, i.e. defining $\Delta X_\alpha$ such that
\begin{align}
    X_\alpha(t_g) = \Delta X_\alpha(b_\alpha^{(A)}\hat{\sigma}_z^{(A)}+b_\alpha^{(B)}\hat{\sigma}_z^{(B)}) \,,
\end{align}
which gives Equation~\eqref{eq:X_ConditionEquations}. An identical procedure gives the expression for $\Delta Y_\alpha$ in Equation~\eqref{eq:Y_ConditionEquations} of the main text.

\subsection{State-dependent phase accumulation during gate operation}
In this section, we derive an expression for the two-qubit entangling phase $\Theta \,\hat{\sigma}_z\otimes \hat{\sigma}_z$ using the action-phase formalism, where the phase acquired by a particular 2Q basis state ($\{\ket{\uparrow\uparrow},\ket{\uparrow\downarrow},\ket{\downarrow\uparrow},\ket{\downarrow\downarrow}\}$) is given by the action associated with the semiclassical trajectory of the atomic wavepacket. 

Treating the position quadratures of each motional mode semiclassically -- i.e. replacing $\hat{X}_\alpha$ with its expectation value $X_\alpha$ -- the Lagrangian for the two-ion system is given by $\mathcal{L} = \sum_{\alpha={\rm CM,SR}}\mathcal{L}_\alpha$, where
\begin{align}  
\label{eq:Mode_Lagrangian}
\mathcal{L}_\alpha(X_\alpha,\dot{X}_\alpha,t) = \frac{m }{2}\left( \dot{X}_\alpha^2 - \lambda_\alpha(t) X^2 \right)l_{0,\alpha}^2 
\end{align}
is the Lagrangian for the $\alpha$-th mode, with lengthscale $l_{0,\alpha}^2=\hbar/(m\omega_\alpha)$. As the motional modes evolve independently from one another, we can consider the phase contribution from each mode independently.

To proceed we define $\phi_\alpha(t_i,t_f)$ as the phase accumulated by the $\alpha-$th mode over the time interval $t\in[t_i,t_f]$, i.e.
\begin{align}
\phi_\alpha(t_i,t_f) \equiv \int_{t_i}^{t_f} dt \;\mathcal{L}_\alpha(X_\alpha,\dot{X}_\alpha,t) \,.
\end{align}
Then, integrating by parts and using the Mathieu-Hill equation $\ddot{X}_\alpha=-\lambda_\alpha(t) X_\alpha$, this can be re-expressed as:
\begin{align}
    \phi_\alpha(t_i,t_f) = \frac{1}{2\omega_\alpha}\left[X_\alpha(t) \dot{X}_\alpha(t) \right]_{t_i}^{t_f} \,.
\end{align}

We can then use this result to find the phase accumulated by the $\mathcal{N}$ SDKs arriving at times $\{t_1,t_2,\dots, t_\mathcal{N}\}$. The motional dynamics are treated piece-wise, with each kick defining a new initial condition for the following interval of free evolution. Specifically, we are interested in the difference in phase acquisition between the phase accumulated during a gate operation ($\phi_{\rm G}$) and the phase accumulated during free evolution ($\phi_{\rm free}$). The former is given by integrating the Lagrangian piece-wise over the intervals $[t_j^+,t_{j+1}^-]$, i.e.
\begin{align}
   \phi_{{\rm G},\alpha } =   \sum_{j=1}^\mathcal{N} \phi_\alpha(t_j^+,t_{j+1}^-) \,,
\end{align}
where we have adopted the notation $t_j^\pm = t_j\pm \epsilon$ where $\epsilon>0$ is an infinitesimal quantity to describe the time immediately before ($t_j^-$) or after ($t_j^+$) the $j$-th SDK. We interpret $t_{\mathcal{N}+1}^-$ to be an arbitrary point in time after the completion of the gate operation, which defines the end of the time interval for the case of free evolution (in the absence of a gate operation):
\begin{align}
    \phi_{{\rm free},\alpha } = \phi_\alpha(t_1^+,t_{\mathcal{N}+1}^-)\,.
\end{align}

Combining the above results, we can express the phase accumulation during a gate operation (relative to the phase accumulated in the absence of a gate) as:
\begin{align}
    \phi_\alpha = \phi_{{\rm G},\alpha }-\phi_{{\rm free},\alpha }  
    = \frac{1}{2} \sum_{j=1}^\mathcal{N} X_\alpha(t_j) \delta Y_{\alpha,j}\,.
\end{align}
Here we have defined $\delta Y_{\alpha,j}$ as the change in the (dimensionless) momentum quadrature for the $\alpha$-th mode due to the $j$-th kick:
\begin{align}
   \delta Y_{\alpha,j}&\equiv  \frac{\dot{X}_\alpha(t_j^+) - \dot{X}_\alpha(t_j^-)}{\omega_\alpha} \\ \notag &=  2^{3/2}z_j \eta_\alpha (b_\alpha^{(A)}\hat{\sigma}_z^{(A)}+b_\alpha^{(B)}\hat{\sigma}_z^{(B)}) \,.
\end{align}
Then, substituting Equation~\eqref{eq:appendix:Xalpha_solution} gives the result:
\begin{widetext}
    \begin{align}
    \phi_\alpha = 8\eta_\alpha^2 \hat{\sigma}_z^{(A)}\otimes \hat{\sigma}_z^{(B)}b_\alpha^{(A)}b_\alpha^{(B)}\sum_{n=1}^\mathcal{N}\sum_{m=2}^{n-1}\frac{z_nz_m}{\rho(t_m)}\left[\sin(\omega_\alpha\Delta t_{nm})\mu_\alpha^{(c)}(t_n,t_m) +\cos(\omega_\alpha\Delta t_{nm})\mu_\alpha^{(s)}(t_n,t_m)\right]
\end{align}
\end{widetext}
where have discarded constant (non-operator-valued) terms which only contribute to an unphysical global phase. Then defining $\Theta$ as
\begin{align}
    \Theta\, \hat{\sigma}_z^{(A)}\otimes \hat{\sigma}_z^{(B)} \equiv \sum_\alpha \phi_\alpha \,,
\end{align}
we arrive at Equation \eqref{eq:Phase_ConditionEquations} of the main text.



\section{Choice of trapping parameters}
\label{app:opti-params}
In this Appendix, we outline how the trap parameters were determined to sensibly compare different micromotion amplitudes ($q_x$ values) in this study. We chose to vary the axial trapping strength to ensure that the system dynamics share the same secular limit across all $q_x$ values considered in this study. As we have reported gate times in units of the radial centre-of-mass frequency, $\omega_0\equiv \omega_{\rm CM}$, it is sufficient to fix the relative splitting between the modes $\chi=\omega_{\rm{SR}}/\omega_{\rm{CM}}-1$.

Our procedure is to choose the equilibrium between two ions, $d$, such that $\chi\approx-1.4\times 10^{-2}$ for each choice of $q_x$. We found that the lowest-order approximation $\beta_\alpha = \sqrt{a_\alpha + q_x^2/2}$ was insufficient for this procedure, particularly for $q_x\gtrsim 0.1$. Instead, the Mathieu characteristic exponents were found numerically through the \textsc{MathieuCharacteristicA} function in Mathematica.

\bibliography{bib}

\end{document}